\documentclass[longauth]{aa}  
\usepackage{xcolor}
\usepackage{support-caption}
\usepackage{subcaption}
\usepackage[autopunct=true]{csquotes}
\usepackage{float}

\usepackage{graphicx}
\usepackage[flushleft]{threeparttable}

\newcommand{\oiiidoub}{[\textrm{O}~\textsc{III}]\ensuremath{\lambda\lambda4959,5007}}
 
\newcommand{\ha}{\ifmmode {\rm H}\alpha \else H$\alpha$\fi}
\newcommand{\hb}{\ifmmode {\rm H}\beta \else H$\beta$\fi}
\newcommand{\lya}{\ifmmode {\rm Ly}\alpha \else Ly$\alpha$\fi}
\newcommand{\pg}{\ifmmode {\rm P}\gamma \else Pa$\gamma$\fi}
\newcommand{\lyb}{\ifmmode {\rm Ly}\beta \else Ly$\beta$\fi}
\newcommand{\lyg}{\ifmmode {\rm Ly}\gamma \else Ly$\gamma$\fi}
\newcommand{\ciii}{\textrm{C}\textsc{III}]\ensuremath{\lambda1908}}

\newcommand{\ciiired}{\textrm{C}\textsc{III}]\ensuremath{\lambda1909}}

\newcommand{\neiii}{[\textrm{Ne}\textsc{III}]\ensuremath{\lambda3869}}

\newcommand{\civ}{\textrm{C}\textsc{IV}\ensuremath{\lambda1548,1550}}

\newcommand{\flyc}{\ifmmode \mathrm{f}_\mathrm{esc}\mathrm{(LyC)} \else $\mathrm{f}_\mathrm{esc}\mathrm{(LyC)}$\fi}

\def\kmsmpc{km s$^{-1}$ Mpc$^{-1}$}

\def\ergs{\ifmmode \mathrm{erg\hspace{1mm}s}^{-1} \else erg s$^{-1}$\fi}

\def\micron{\ifmmode \mu\mathrm{m} \else $\mu$m\fi}
\def\msun{\ifmmode \mathrm{M}_{\odot} \else M$_{\odot}$\fi}
\def\msunyr{\ifmmode \mathrm{M}_{\odot} \hspace{1mm}{\rm yr}^{-1} \else $\mathrm{M}_{\odot}$ yr$^{-1}$\fi}
\def\zsun{\ifmmode Z_{\odot} \else Z$_{\odot}$\fi}
\def\lsun{\ifmmode L_{\odot} \else L$_{\odot}$\fi}
\def\mstar{\ifmmode \mathrm{M}_{\star} \else M$_{\star}$\fi}

\newcommand{\jwst}{JWST}

\newcommand{\MIRI}

\usepackage{tabularx}
\usepackage{lipsum} 
\usepackage{array} 
\usepackage{amsmath} 

\usepackage{txfonts}
\newcommand{\orcid}[1]{\href{https://orcid.org/#1}{\includegraphics[width=10pt]{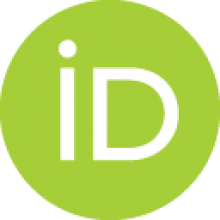}}}
\usepackage{hyperref}
\hypersetup{colorlinks, 
 linkcolor = blue,
 urlcolor = blue,
 citecolor = blue,
 anchorcolor = blue
}
\begin{document}


\title{Seven wonders of Cosmic Dawn: JWST confirms a high abundance of galaxies and AGNs at z $\simeq$ 9-11 in the GLASS field}
\titlerunning{Spectroscopic confirmation of z $\simeq$10 galaxies in the GLASS-JWST ERS field}
\authorrunning{L. Napolitano et al.}



 \subtitle{}
     \author{L. Napolitano \orcid{0000-0002-8951-4408}
 \inst{1,2}
 \and M. Castellano \orcid{0000-0001-9875-8263}
 \inst{1}
 \and L. Pentericci \orcid{0000-0001-8940-6768}
 \inst{1}
 \and P. Arrabal Haro \orcid{0000-0002-7959-8783}
 \inst{3}
 \and A. Fontana \orcid{0000-0003-3820-2823}
 \inst{1}
 \and T. Treu \orcid{0000-0002-8460-0390}
 \inst{12}
 \and \\ P. Bergamini \orcid{0000-0003-1383-9414}
 \inst{4,5}
 \and A. Calabrò \orcid{0000-0003-2536-1614}
 \inst{1}
 \and S. Mascia \orcid{0000-0002-9572-7813}
 \inst{1,8}
 \and T. Morishita \orcid{0000-0002-8512-1404}
 \inst{9}
 \and G. Roberts-Borsani \orcid{0000-0002-4140-1367}
 \inst{11}
 \and P. Santini \orcid{0000-0002-9334-8705}
 \inst{1}
 \and \\ E. Vanzella \orcid{0000-0002-5057-135X}
 \inst{13}
 \and B. Vulcani \orcid{0000-0003-0980-1499}
 \inst{6}
 \and D. Zakharova \orcid{0009-0001-1809-4821}
 \inst{6,7}
 \and T. Bakx \orcid{0000-0002-5268-2221}
 \inst{19}
 \and M. Dickinson \orcid{0000-0001-5414-5131}
 \inst{3}
 \and C. Grillo \orcid{0000-0002-5926-7143}
 \inst{4} 
 \and \\ N. Leethochawalit \orcid{0000-0003-4570-3159}
 \inst{10}
 \and M. Llerena \orcid{0000-0003-1354-4296}
 \inst{1} 
 \and E. Merlin \orcid{0000-0001-6870-8900}
 \inst{1}
 \and D. Paris \orcid{0000-0002-7409-8114}
 \inst{1}
 \and S. Rojas-Ruiz \orcid{0000-0003-2349-9310}
 \inst{12}
 \and P. Rosati \orcid{0000-0002-6813-0632}
 \inst{5,13}
 \and \\ X. Wang \orcid{0000-0002-9373-3865}
 \inst{16,17,18}
 \and I. Yoon \orcid{0000-0001-9163-0064}
 \inst{14}
 \and J. Zavala \orcid{0000-0002-7051-1100}
 \inst{15}
  %
  %
 }
 \institute{\textit{INAF – Osservatorio Astronomico di Roma, via Frascati 33, 00078, Monteporzio Catone, Italy}\\ 
 \email{lorenzo.napolitano@inaf.it}
 \and 
 \textit{Dipartimento di Fisica, Università di Roma Sapienza, Città Universitaria di Roma - Sapienza, Piazzale Aldo Moro, 2, 00185, Roma, Italy} 
 \and
 \textit{NSF’s National Optical-Infrared Astronomy Research Laboratory, 950 N. Cherry Ave., Tucson, AZ 85719, USA} 
 \and
 \textit{Dipartimento di Fisica, Università degli Studi di Milano, Via Celoria 16, I-20133 Milano, Italy} 
 \and
 \textit{Osservatorio di Astrofisica e Scienza dello Spazio di Bologna, via Gobetti 93/3, I-40129 Bologna, Italy} 
 \and 
 \textit{INAF - Osservatorio astronomico di Padova, Vicolo Osservatorio 5, 35122, Padova, Italy} 
 \and
 \textit{Dipartimento di Fisica e Astronomia 'Galileo Galilei', Universita' degli studi di Padova, Vicolo dell'Osservatorio, 3, I-35122 Padova, Italy} 
 \and 
 \textit{Dipartimento di Fisica, Università di Roma Tor Vergata, Via della Ricerca Scientifica, 1, 00133, Roma, Italy} 
 \and
 \textit{IPAC, California Institute of Technology, MC 314-6, 1200 E. California Boulevard, Pasadena, CA 91125, USA} 
 \and
 \textit{National Astronomical Research Institute of Thailand (NARIT), Mae Rim, Chiang Mai, 50180, Thailand} 
 \and
 \textit{Department of Astronomy, University of Geneva, Chemin Pegasi 51, 1290 Versoix, Switzerland} 
 \and
 \textit{Department of Physics and Astronomy, University of California, Los Angeles, 430 Portola Plaza, Los Angeles, CA 90095, USA} 
 \and
 \textit{Dipartimento di Fisica e Scienze della Terra, Universit`a degli Studi di Ferrara, Via Saragat 1, I-44122 Ferrara, Italy} 
 \and
 \textit{National Radio Astronomy Observatory, 520 Edgemont Road, Charlottesville, VA 22903, USA} 
 \and
 \textit{National Astronomical Observatory of Japan, 2-21-1, Osawa, Mitaka, Tokyo, Japan} 
 \and
 \textit{School of Astronomy and Space Science, University of Chinese Academy of Sciences (UCAS), Beijing 100049, China} 
 \and
 \textit{National Astronomical Observatories, Chinese Academy of Sciences, Beijing 100101, China} 
 \and
 \textit{Institute for Frontiers in Astronomy and Astrophysics, Beijing Normal University, Beijing 102206, China} 
 \and
 \textit{Department of Space, Earth, \& Environment, Chalmers University of Technology, Chalmersplatsen 4 412 96 Gothenburg, Sweden} 
 %
}

\date{Accepted XXX. Received YYY; in original form ZZZ}
 
\abstract{We present \jwst/NIRSpec PRISM follow-up of candidate galaxies at z$\simeq$ 9-11 selected from deep \jwst/NIRCam photometry in GLASS-JWST Early Release Science data. 
We spectroscopically confirm six sources with secure redshifts at z = 9.52–10.43, each showing multiple emission lines. An additional object is likely at z $\simeq$ 10.66, based on its \lya-break and a single emission feature, while one source is a lower redshift interloper.
The sample includes the first JWST-detected candidate at z$\sim$10, GHZ1/GLASS-z10, which we confirm at z = 9.875, and the X-ray detected AGN GHZ9 confirmed at z = 10.145. Three objects in our sample, including GHZ9, have EW(\ciii)>20\AA\ and occupy a region compatible with AGN emission in the EW(CIII]) vs CIV/CIII] diagram. The spectroscopic sample confirms a high abundance of galaxies at z $>$ 9. We measure a number density of z$\sim$10 galaxies in the GLASS-JWST ERS field that is a factor of $>$3 higher than other JWST-based estimates at demagnified rest-frame magnitudes of -21 $\leq M_{\mathrm{UV}} \leq$ -19. We find that the positions of these galaxies in redshift and angular space are not consistent with all of them being part of a unique progenitor of present-day galaxy clusters.
The high density of objects in the GLASS region can be explained either by clustering on large scales or by a superposition of different forming structures of which we observe only the brightest members. By considering all the spectroscopic z$\sim$10 sources in the Abell-2744 field, we identify two potential galaxy proto-clusters centered around GHZ9 and JD1, with relative separations between their members of $\simeq$1-2 pMpc. The potential AGN nature of three of the sources in our sample lends support to a scenario in which the high abundance of bright sources determined by JWST surveys at cosmic dawn may be affected by AGN contribution to their UV luminosity.}
 



 \maketitle

 


\section{Introduction} \label{sec:intro}
The investigation of galaxy populations in the first few hundred million years of the Universe’s lifetime is a primary goal in the extragalactic field. Galaxies and overdensities emerging from this early epoch are the seeds of present-day objects and galaxy clusters. Observing the first generation of galaxies at Cosmic Dawn, covering more than 13 billion years of evolution from the present-day Universe, provides a crucial test for galaxy evolution models.
Deep observations with the Hubble Space Telescope (HST) and the Spitzer Space Telescope have provided the identification of galaxies up to $z$ $\sim$ 10 \citep[e.g.,][]{Bouwens2015, Oesch2016, Finkelstein2022a, Roberts-Borsani2022}, charting the evolution of their physical properties.
The goal of identifying galaxies at even higher redshifts is currently being advanced by the \textit{James Webb} Space Telescope \citep[\jwst ;][]{Gardner2006, Gardner2023}, which is further extending our observational horizon. This expansion includes the capability to obtain photometry up to 5 $\mu$m using the NIRCam \citep[][]{Rieke2005, Rieke2023NIRCam} instrument, effectively probing galaxy candidates at $z$ > 10 \citep[e.g.,][]{Castellano2022b, Finkelstein2022b, Naidu2022b, Donnan2023, Robertson2023}. 

The results gathered in the first two years of JWST operations have revealed a surprisingly high abundance of bright high-redshift galaxies compared to the predictions of theoretical models and extrapolations from lower-redshift estimates \citep[e.g.,][]{Castellano2023, Finkelstein2023, Finkelstein2023b, Harikane2023, PerezGonzalez2023}. 
Understanding the origin of this \textquote{excess} of bright z $\gtrsim$ 9 galaxies is among the most pressing goals of extragalactic astrophysics. In fact, the observations suggest that we may be missing key astrophysical processes that shape the formation of the first galaxies, such as a higher star-formation efficiency at early times \citep[e.g.,][]{Harikane2023, Harikane2024, Mason2023}, an increased luminosity due to a lower dust extinction \citep[e.g.,][]{Ferrara2023, Ferrara2024, Tsuna2023, Ziparo2023}, the contribution of active galactic nuclei \citep[AGNs, e.g.,][]{Pacucci2022}, or of metal poor stars with a top-heavy IMF \citep[e.g.,][]{Haslbauer2022, Inayoshi2022, Trinca2024, Yung2024}. Also, non-standard cosmologies have been investigated to reconcile theoretical predictions with observations \citep[e.g.,][]{Menci2022, Menci2024, Melia2023, Padmanabhan2023, McGaugh2024}. 

After the initial discovery from deep imaging surveys, \jwst\ spectroscopic follow-up observations have successfully confirmed the high number density of $z$ $\sim$ 10 galaxies \citep[e.g.,][]{ArrabalHaro2023, Bunker2023B, Curtis-Lake2023, Roberts-Borsani2023, Carniani2024, Castellano2024, Harikane2024, Hsiao2024, Witstok2024}, finding very few cases of interlopers and redshift misidentifications in the original photometric samples \citep[e.g.,][]{Arrabal_Haro2023Nature, Zavala2023, Harikane2024, Harikane2024B}.
However, several ingredients are missing to fully understand the origin of the abundant population of bright high-redshift galaxies. On the one hand, unprecedented efforts are being spent on the physical characterization of these early objects using deep NIRSpec and MIRI spectroscopy to constrain their metallicity, dust extinction, star-formation history, and AGN contribution \citep[e.g.,][]{Bunker2023B,Cameron2023,Maiolino2023,Hsiao2024, Roberts-Borsani2024,Castellano2024,Calabro2024B,Zavala2024}. On the other hand, a crucial aspect that spectroscopic follow-up campaigns have not fully addressed yet is the role of clustering and cosmic variance on UV LF estimates at such high-redshifts. Understanding galaxy clustering in the epoch of cosmic dawn is fundamental both to achieve firm constraints on the slow evolution of the \textit{average} number density of bright galaxies at z $\gtrsim$ 9, and to understand their role in the earliest phases of cosmic reionization. In fact, the analysis of galaxy populations at z $\sim$ 6-8 suggests that reionization was a spatially inhomogeneous process \citep[][]{Treu2012,Pentericci2014}. Consistently with \textquote{inside-out} reionization scenarios, significant differences have been found in the abundance of galaxies in different lines-of-sight due to the presence of proto-clusters associated to the first \textquote{reionized bubbles} \citep[][]{Castellano2016, Tilvi2020, Jung2021, Leonova2021, Endsley2022}. 
Spectroscopic observations conducted with the \jwst\ NIRSpec have demonstrated its capability to successfully identify proto-clusters \citep[e.g.,][]{Morishita2022,Arribas2024} and filament-like structures \citep[e.g.,][]{Chen2024, Napolitano2024} at z $\geq$ 7. 

The ideal testing ground to explore the role of clustering at the onset of reionization is provided by the GLASS-JWST ERS field \citep[][]{Treu2022} where a significant excess of z$\sim$10 candidates was found compared to other JWST fields \citep[][C23 hereafter]{Castellano2023}. The first set of GLASS-JWST NIRCam observations led to the discovery of GHZ1/GLASS-z10, the first bright JWST-selected candidate at z$\sim$10 \citep[][]{Castellano2022b,Naidu2022b}\footnote{The two discovery papers appeared on arXiv the same day, and named the galaxy GHZ1 \citep{Castellano2022b} and GLASS-z10 \citep{Naidu2022b}. In the remainder of the paper, we refer to it only as GHZ1 for conciseness.}. The analysis of the final GLASS-JWST NIRCam imaging by C23 identified the presence of a significant overdensity of z$\sim$9-11 candidates in the region \citep[see also][]{McLeod2024,Chemerynska2024}, possibly extending through $\sim2$ Mpc in projected length to the surrounding UNCOVER and DDT\#2756 regions.
In this paper, we present the results of NIRSpec follow-up of the z $\sim$ 10 candidates in the GLASS-JWST ERS field obtained through Program GO-3073 (PI M. Castellano). The multi-object spectroscopy observations targeted all five z $\sim$ 9--11 Lyman-break galaxies from C23, including GHZ1 and the candidate z$\sim$10 AGN GHZ9 \citep[][]{Kovacs2024}, plus additional targets in the same redshift range from \cite{Castellano2022b}, \cite{Harikane2023}, \cite{Atek2023b}, and \cite{McLeod2024}. The main goal of the present work is to constrain the number density of z$\sim$10 galaxies and assess the presence of an overdensity in the GLASS field in the light of the confirmed spectroscopic redshifts. A more detailed assessment of the physical properties of the objects is deferred to a forthcoming work. 

The paper is organized as follows. We describe observations and data reduction in Sect.~\ref{sec:Data}, and the adopted methodology in Sect.~\ref{sec:Methods}. The spectroscopically-confirmed galaxies are discussed in Sect.~\ref{sec:confirmations}. We discuss the implications on the UV luminosity function and the possible enhanced  clustering in the field in Sect.~\ref{sec:abundance}. Our findings are summarised and discussed in Sect.~\ref{sec:Summary}. In the following, we adopt the $\Lambda$CDM concordance cosmological model ($H_0 = 70$ \kmsmpc, $\Omega_M = 0.3$, and $\Omega_{\Lambda} = 0.7$). We report all magnitudes in the AB system \citep{Oke1983} and equivalent widths (EW) to rest-frame values.

\section{Observations and data reduction} \label{sec:Data}

The \jwst\ NIRSpec \citep[][]{Jakobsen2022} PRISM data employed in this work were taken as part of the Cycle 2 program GO-3073 (PI M. Castellano). The program is designed to target the GLASS-JWST NIRCam field \citep[][]{TreuGlass2022} to obtain  spectroscopic observations of the $z \geq$ 10 candidates selected by \cite{Castellano2022b} and \cite{Castellano2023}, by observing two partially overlapping pointings in two epochs with three configurations each. The MSA have been designed in order to reach S/N $\gtrsim$ 5 in the continuum at the \lya-break. We thus observed the brightest candidates from C23 with three configurations each, while the faintest ones (GHZ4 and GHZ9) have been included in both epochs. Additional high-redshift candidates have been included with high priority whenever observable. In this paper, we consider both data obtained during the first epoch (2023 Oct. 24) and spectra from the second pointing (2024 Jul. 5). We adopt a standard three-point nodding pattern in NRSIRS2 readout mode on three-shutter long slits for all the multi-shutter array \citep[MSA,][]{Ferruit2022} primary targets in the NIRSpec Multiobject Spectroscopy (MOS) configuration. 
This strategy facilitates background subtraction. The observation of each pointing is further divided into three visits, with an exposure time of 6567 s each. Unfortunately, an electrical short affected the third visit of the first pointing. 
Thus, we exploit 6 and 9 dithered observations from the first and second pointings respectively, for a total of 32835 s. \\
The selected instrument configuration provides continuous wavelength coverage in the 0.6-5.3 $\mu$m wavelength range, with varying spectral resolution $R \sim$ 30 - 300. As described by \cite{Curtis-Lake2023}, the PRISM disperser offers a significant advantage for high-redshift spectroscopic identification. Its low resolution at bluer wavelengths enables the detection of faint UV continuum and the \lya-break feature, while the increasing spectral resolution at redder wavelengths allows for robust constraints on optical line detection.

\subsection{Imaging data and target selection} \label{sec:Nircam}
The galaxies discussed in this paper were selected as $z$ > 9 candidates from the GLASS-JWST Early Release Science \citep[JWST-ERS-1324, PI T. Treu, see][]{TreuGlass2022} program leveraging NIRCam photometry in the F090W, F115W, F150W, F200W, F277W, F356W, and F444W bands \citep[see,][]{Paris2023}. The southernmost part of the field was also observed by the UNCOVER program \citep[GO-2561, PIs Labbé and Bezanson, see][]{Bezanson2022} including observations in the F410M filter. In this work we use an updated measurement of the NIRCam photometry in the A2744 field which exploits the latest reduction of the GLASS-JWST NIRCam data including new observations acquired in July 2023. The updated multi-band photometry is discussed in detail in Merlin et al., (subm). Briefly, sources were detected in a weighted average image of the F356W and F444W bands using \textsc{SExtractor} \citep[][]{Bertin1996}. Flux measurements were performed with \textsc{A-PHOT} \citep{Merlin2019}. The total flux of each source was measured on the detection image within \citet{Kron1980} apertures, while fluxes in the other bands were derived by scaling the aforementioned total flux according to colors obtained on point-spread-function (PSF) matched images within an aperture with a diameter twice the PSF full width at half-maximum intensity \citep[FWHM = $0\farcs28$, see also][]{Merlin2022,Paris2023}. 
 
We designed the MSA masks to follow-up all the five color-selected z$\sim$10 candidates within the GLASS-JWST ERS region discussed in C23, plus object GHZ2, which was confirmed at z = 12.342 by \cite{Castellano2024} \citep[see also,][]{Zavala2024, Calabro2024B}. Whenever possible, additional candidates from \citet{Castellano2022b}, \citet{Atek2023b}, \citet{Harikane2023} and \citet{McLeod2024} were included, as well as objects with photometric redshift z $\gtrsim$ 9 from Merlin et al. (subm) or from \citet{Paris2023}. In total, we observed 27 candidates with a photometric redshift solution $\gtrsim$ 9.
Remaining slitlets were allocated to lower-redshift sources down to the local Universe (Vulcani et al., in prep.) for a total of 978 targets observed by the GO-3073 program.

The GLASS field is affected by magnification from the foreground Abell-2744 cluster at $z$ = 0.3072. In our analysis, we will correct rest-frame quantities for magnification ($\mu$, see Table \ref{tab:summary_data}) on the basis of the model by \cite{Bergamini2023}, which incorporates data from 149 multiple images corresponding to 50 background sources, of which 121 have been spectroscopically-confirmed.

\subsection{NIRSpec data reduction} \label{sec:Nirspec}
\begin{figure*}[ht!]
\begin{minipage}{\textwidth}
\centering
\includegraphics[width=\linewidth]{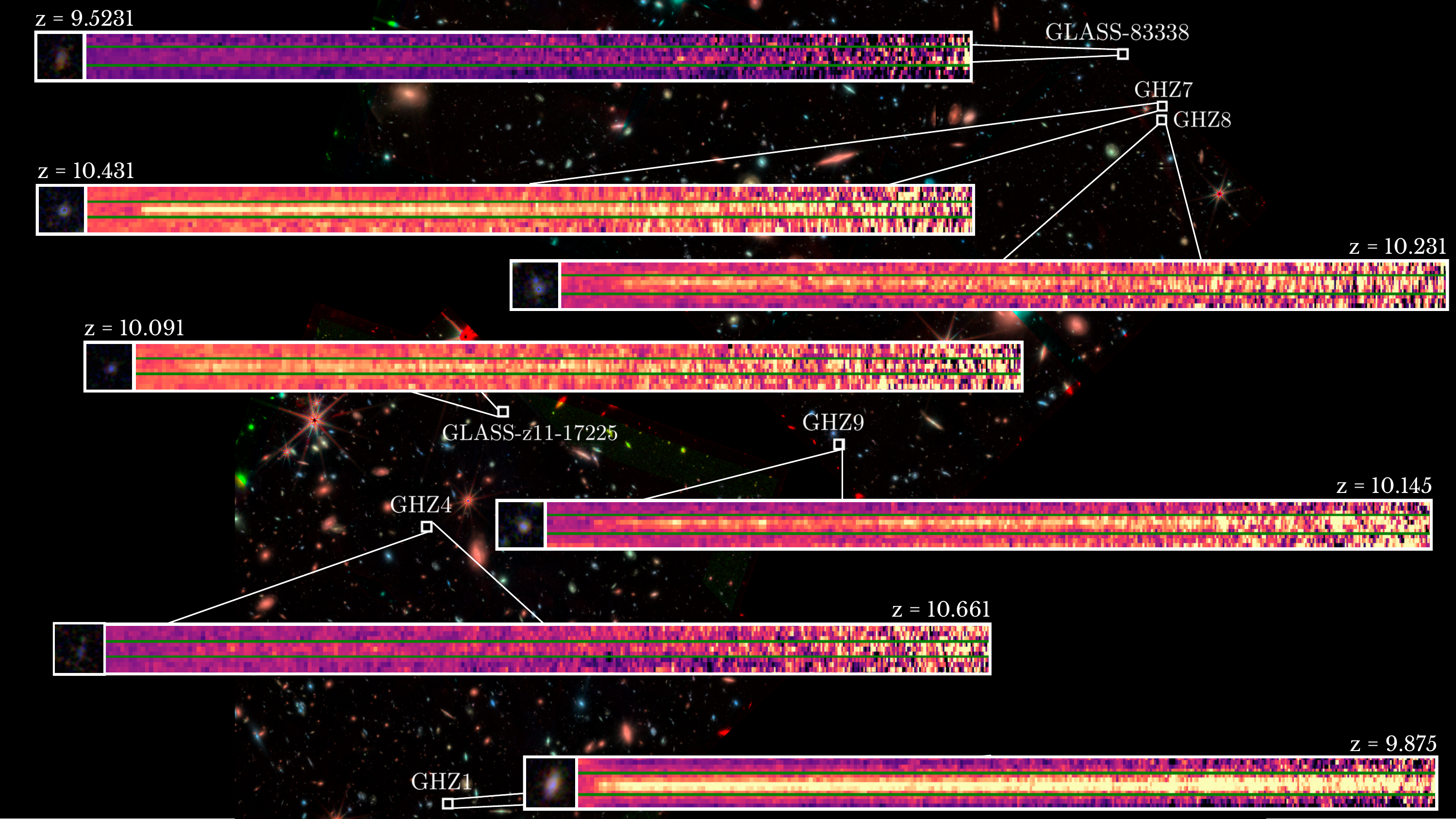}
\end{minipage}
\caption{Spatial distribution of the seven galaxies spectroscopically-confirmed at 9 < z < 11 in the GLASS field, overlaid on a JWST/NIRCam RGB image from the GLASS-JWST program \citep[][red: F200W, green: F356W, blue: F444W]{Paris2023}. For each source, we show the NIRSpec 2D spectrum and a zoom-in image of the galaxy.} \label{fig:pointing}
\end{figure*}

The NIRSpec data were reduced as outlined by \cite{ArrabalHaro2023} and \cite{Arrabal_Haro2023Nature} with the STScI Calibration Pipeline\footnote{https://jwst-pipeline.readthedocs.io/en/latest/index.html} version 1.13.4, and the Calibration Reference Data System (CRDS) mapping 1197. The data reduction starts with reprocessing the individual uncalibrated (uncal) exposures obtained from the Mikulski Archive for Space Telescopes (MAST) using the official JWST Pipeline. At this stage, the \textsc{calwebb\_detector1} module subtracts dark current and bias, and generates count-rate maps (CRMs) from the uncalibrated images. The output is then processed with the \textsc{calwebb\_spec2} module, which creates two-dimensional (2D) cutouts of the slitlets, corrects for flat-fielding, performs background subtraction using the three-nod pattern and slit loss correction, executes photometric and wavelength calibrations, and resamples the 2D spectra to correct distortions of the spectral trace. At this stage we visually inspected all the outputted 2D spectra from the dithered observations of the visits, to search for possible interlopers that would affect the standard three-nod background subtraction of the primary sources. For GLASS-83338, GHZ9, and GHZ4 we found that some dithered observations include a secondary object 
in the slit. We therefore masked the contaminated 2D regions before re-applying the three-nod pattern background subtraction. Next, the \textsc{calwebb\_spec3} module combines the images of the individual nods, utilizing customized extraction apertures to extract the one-dimensional (1D) spectra. We found that this increases the average signal-to-noise ratio (S/N) by up to a factor of 1.25 compared to the standard MSA extraction apertures. Finally, we visually inspected both 2D and 1D spectra using the \textsc{Mosviz} visualization tool \citep{Developers2023} to identify and mask any lingering hot pixels and artifacts in the spectra.
In Fig.~\ref{fig:pointing} we show the spatial distribution of the seven galaxies confirmed to be at $z$ > 9, while in Fig.~\ref{fig:spectra} we present the final 2D and 1D spectra. \\
To investigate potential residual issues in the absolute flux calibration caused by slit losses or other inaccuracies in flux calibration files, we further checked the consistency with the broad band photometry by integrating the spectra across the NIRCam F150W, F200W, F277W, F356W, and F444W filter bandpasses. We only considered the resulting synthetic photometry with S/N > 5 and compared it with the measured NIRCam photometry. From this process, we derived correction factors for the spectra in each NIRCam filter. The average correction factor agrees within its error, with the individual correction factors. Therefore, in our data analysis, we applied the average correction to the fluxes and the associated error values for each galaxy, making the correction independent of wavelength. The resulting multiplicative flux correction factors are consistent with those reported by \cite{Arrabal_Haro2023Nature} and \cite{Napolitano2024}.\\
We note that the photometric correction, as well as the correction for magnification, do not affect the measurement of EWs and line ratios. 

\begin{figure*}[]
    \centering
    \includegraphics[width=0.75\textwidth]{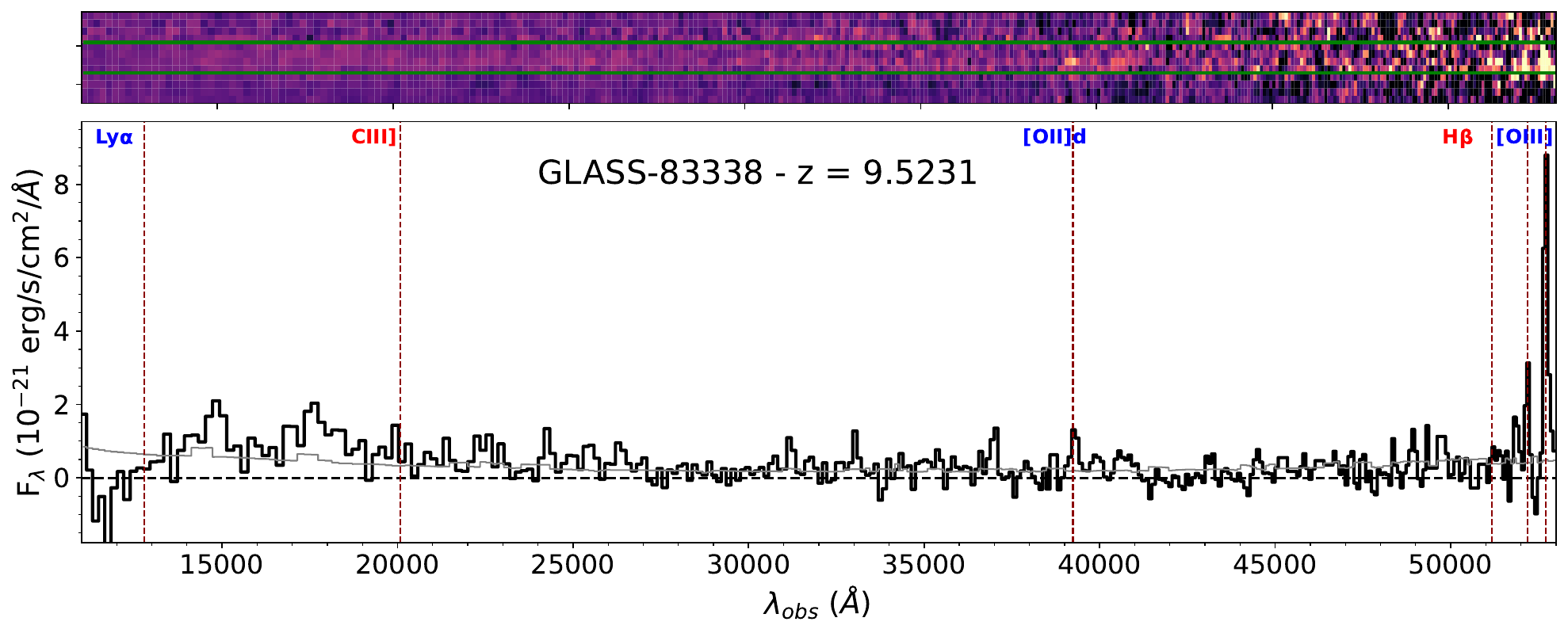}
    \includegraphics[width=0.75\textwidth]{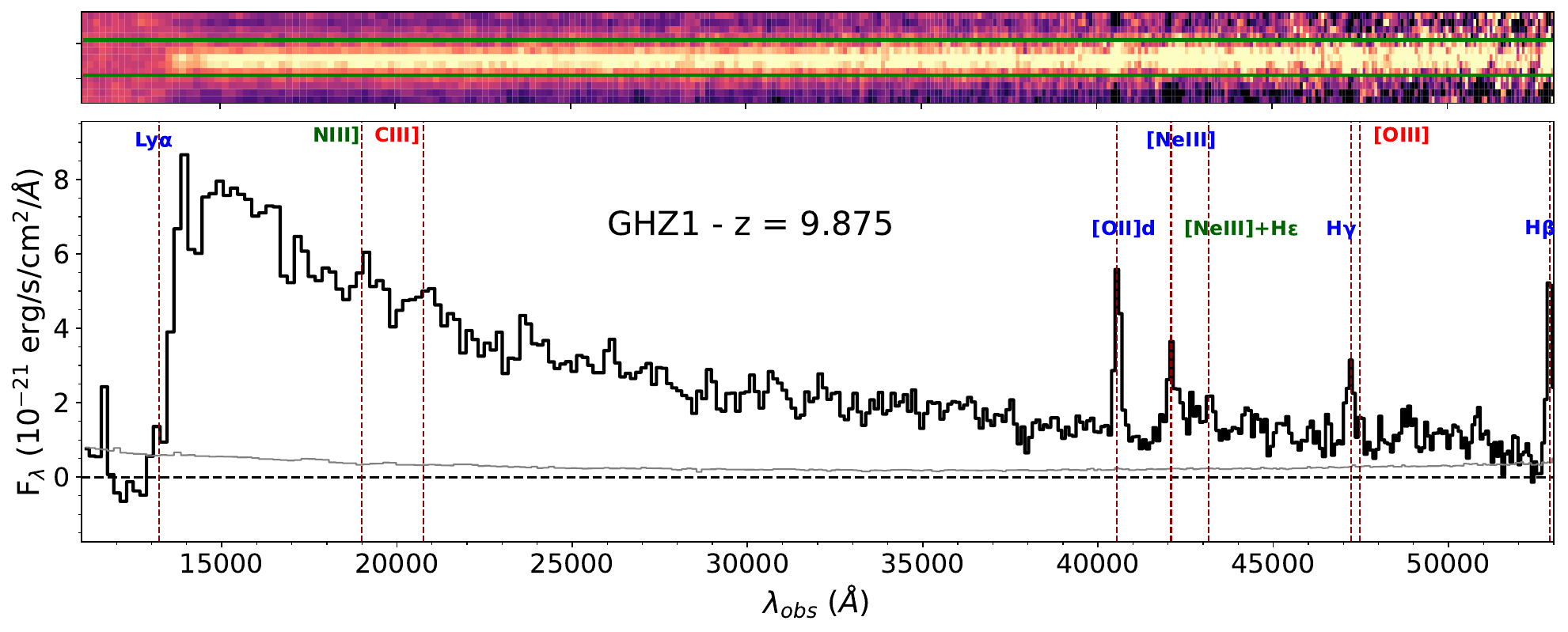}
    \includegraphics[width=0.75\textwidth]{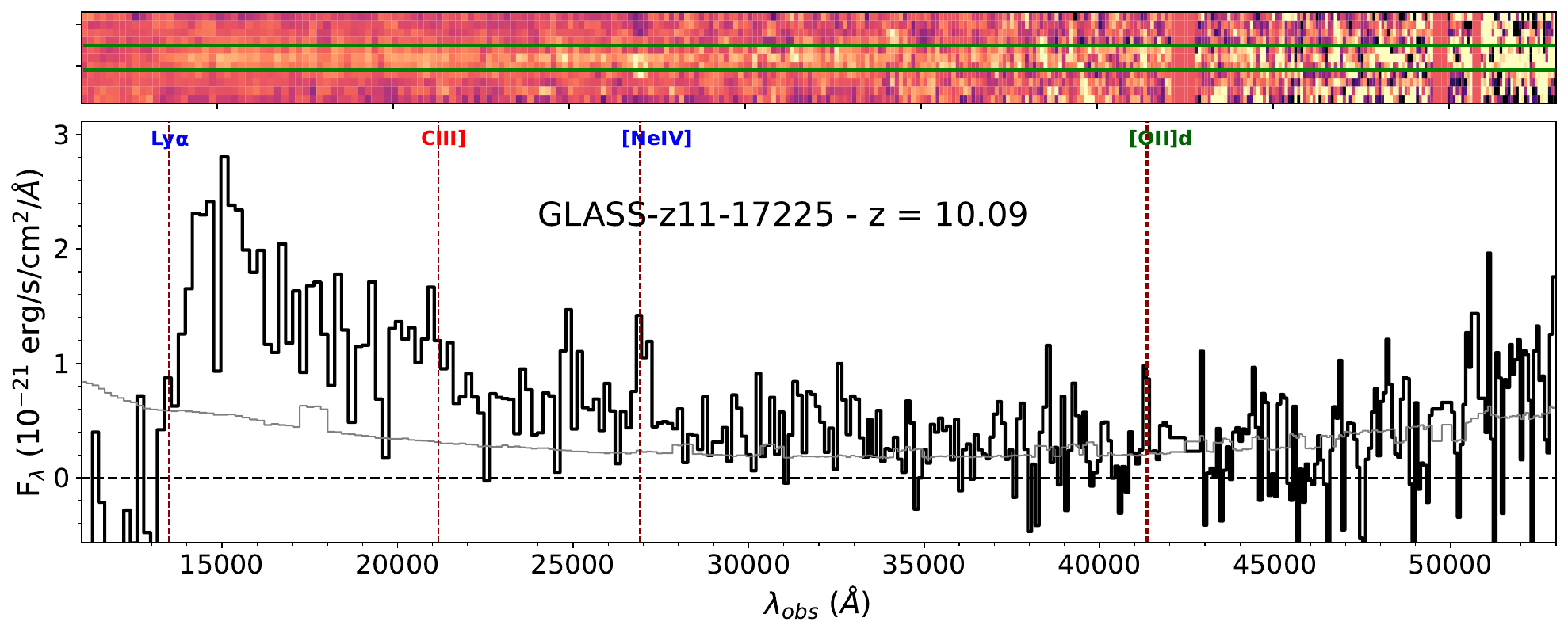}
    \includegraphics[width=0.75\textwidth]{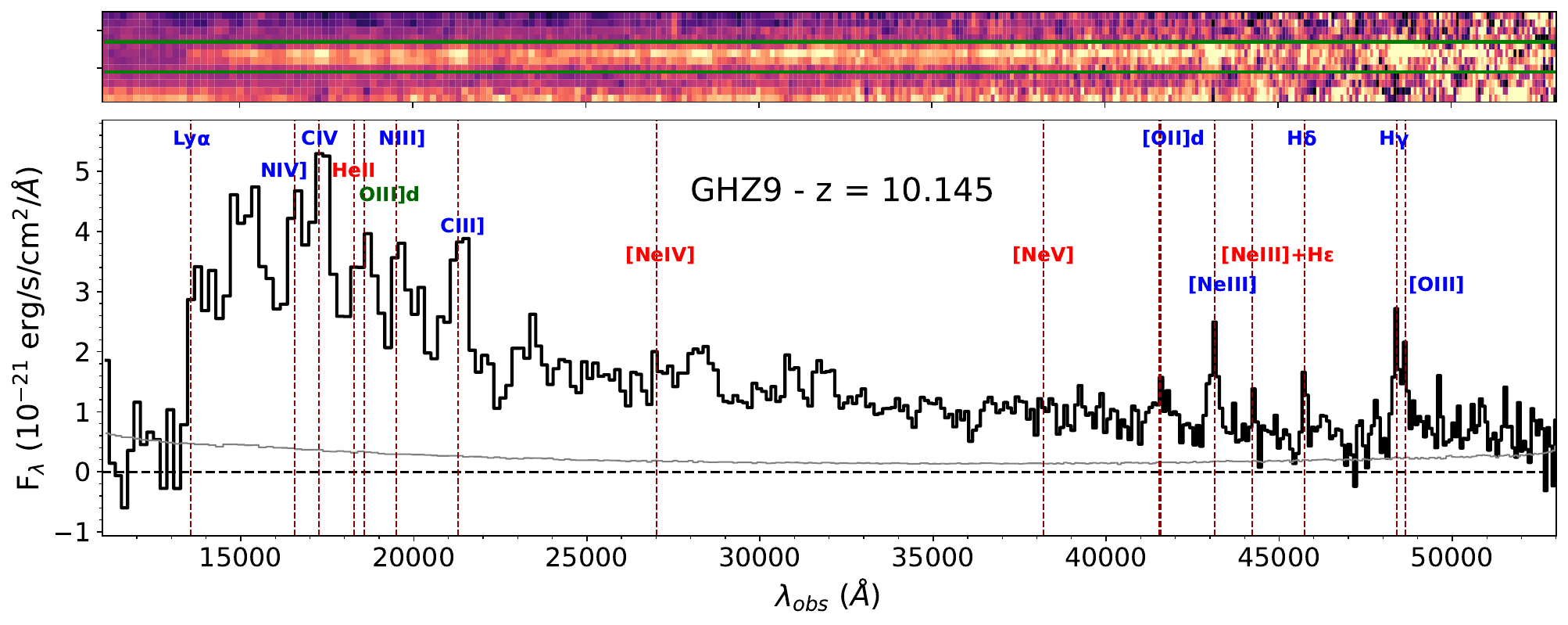}
    \caption{Observed 2D (upper panels) and 1D (lower panels) PRISM spectra of the confirmed z > 9 galaxies from the data reduction. The horizontal green lines enclose the customized extraction regions where we extract the 1D spectra. The pipeline error spectrum is reported in grey. We show emission lines with an integrated S/N > 5 in blue, while we report significant (3 < S/N < 5) and tentative (S/N < 3) emission features in green and red, respectively. The \lya-break feature is always shown.}
    \label{fig:spectra}
\end{figure*}

\begin{figure*}[]
    \ContinuedFloat
    \centering
    \includegraphics[width=0.75\textwidth]{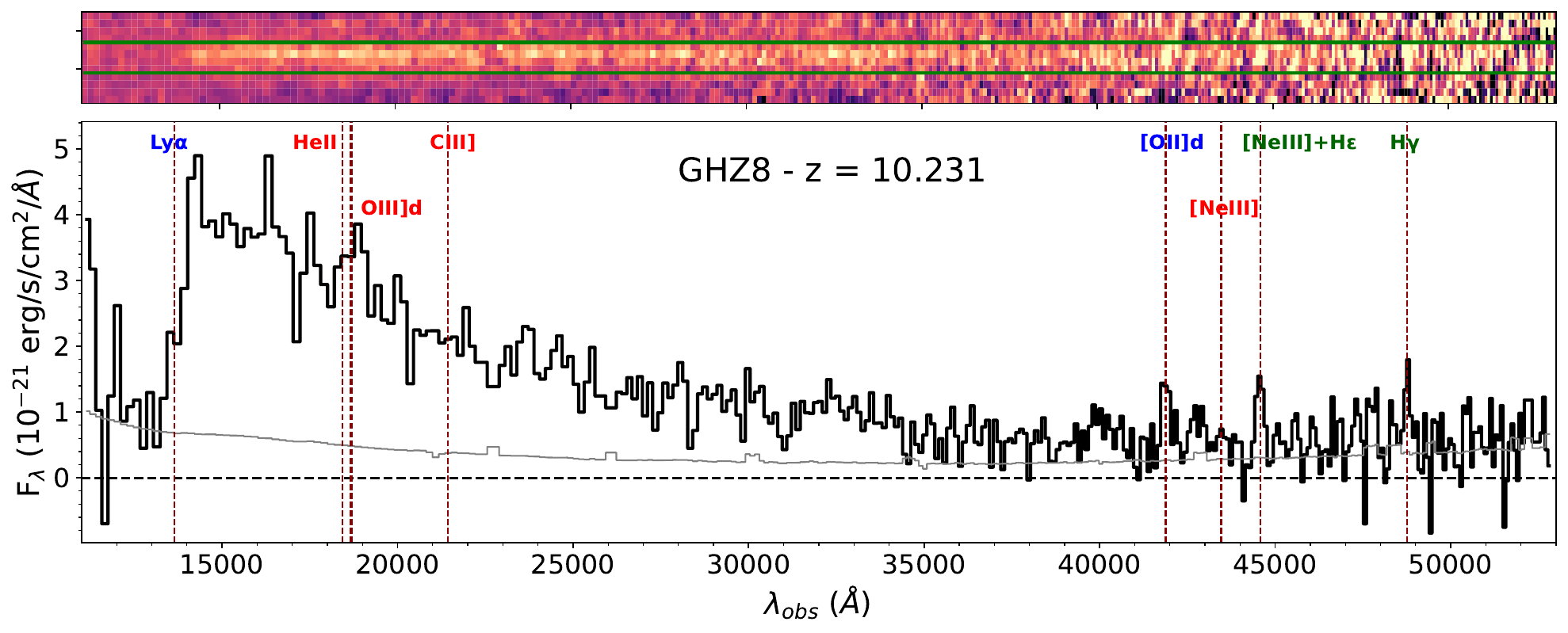}
    \includegraphics[width=0.75\textwidth]{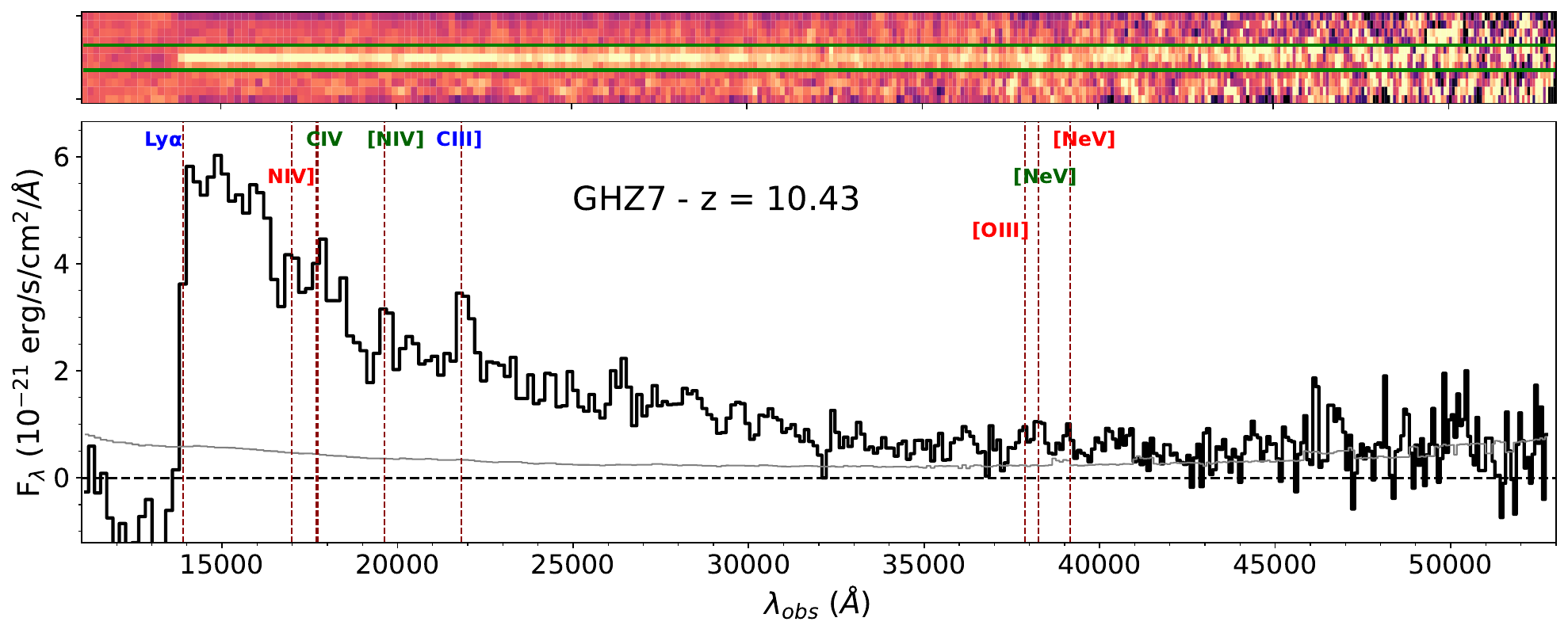}
    \includegraphics[width=0.75\textwidth]{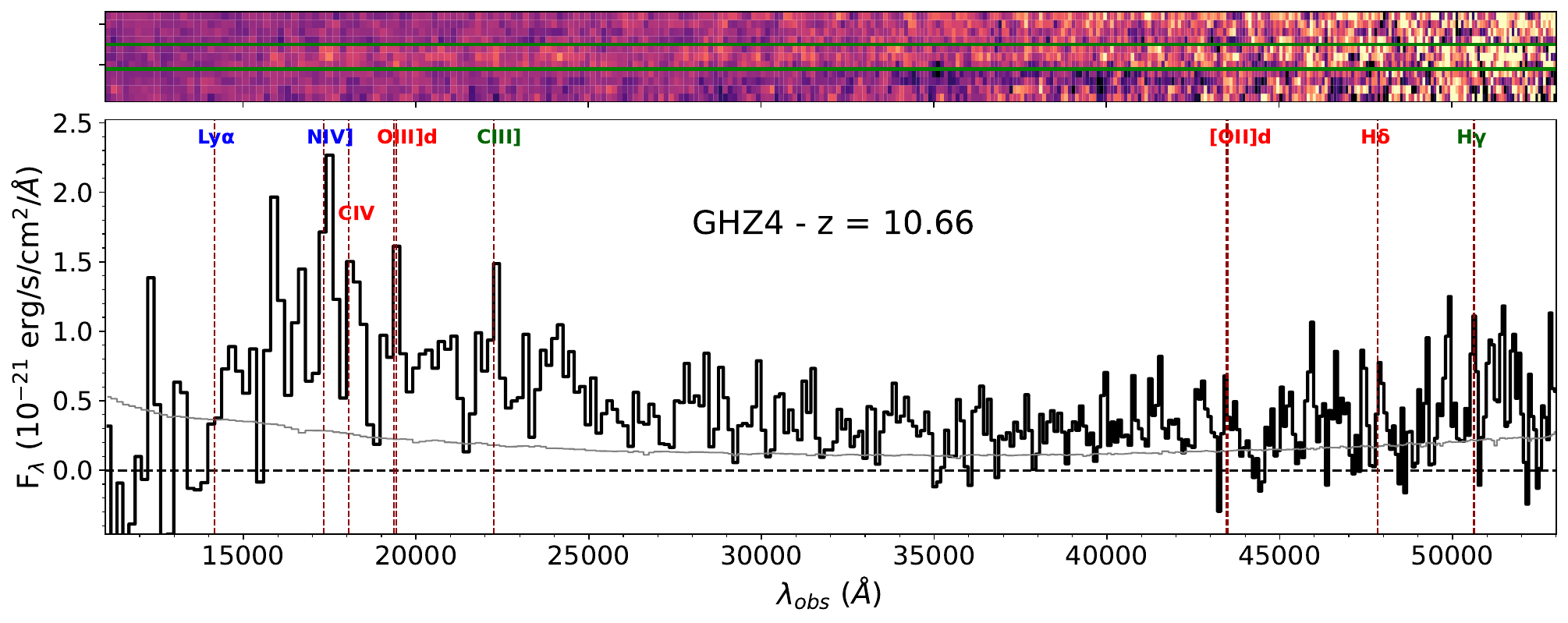}
    \caption{continued.}
    \label{fig:spectra2}
\end{figure*}

\section{Methods} \label{sec:Methods}
\subsection{Spectroscopic redshift determination} \label{sec:fluxspec}

Each spectrum was visually inspected to search for the \lya-break and emission lines features. As the resolution of the PRISM steadily increases with wavelength beyond 1$\mu$m, for an initial redshift estimate of each source, we used the line centroid of the reddest emission line identified.\\
We regard integrated line emissions with a S/N > 3 as significant, with uncertainty accounting for errors in both the integrated line flux and the extrapolated continuum at the line position. The continuum is estimated through a linear interpolation of regions closest to each line that are free from potential features, using the \textsc{emcee} \citep[][]{Foreman_Mackey2013} routine. To determine a first estimate of line fluxes, we directly integrate the continuum-subtracted spectrum within a window centered at the expected observed wavelength and spanning 4 $\times$ $\sigma_R(\lambda)$, where $\sigma_R(\lambda)$ represents the expected Gaussian standard deviation for a line observed at resolution R($\lambda$). The resolution is provided by the \jwst\ documentation\footnote{\url{https://jwst-docs.stsci.edu/jwst-near-infrared-spectrograph/nirspec-instrumentation/nirspec-dispersers-and-filters}} with the assumption of a source that illuminates the slit uniformly. For partially blended lines, we assess the significance of the individual components within narrower windows of 2 $\times$ $\sigma_R(\lambda)$.\\
For each spectrum, after directly integrating all the features identified, we are left with a list of line emissions with S/N > 3 and 3$\sigma$ limits for features with S/N < 3. We recompute the spectroscopic redshift by taking a weighted average of the line centroids from S/N > 5 emission lines only.
As detailed in \cite{Castellano2024}, for all significant emission features, we perform a Gaussian fit on the continuum-subtracted flux using the \textsc{specutils} package of \textsc{astropy} \citep{Astropy2013}. Unresolved doublets and multiplets are fitted as a single Gaussian profile, while partially blended lines are fitted with  a double-Gaussian profile. The mean of the Gaussian is allowed to vary with $\Delta$z = 0.04, and the Gaussian standard deviation within 5\% of the nominal $\sigma_R(\lambda)$, to account for redshift uncertainties. We further refine the Gaussian fit by employing \textsc{emcee} to perform a Markov chain Monte Carlo (MCMC) analysis. The amplitude, standard deviation and their uncertainty obtained from the best Gaussian model outputted by the \textsc{specutils} package are used to initialize 100 walkers with 100,000 iterations. The best model parameters and the integrated flux are determined by taking the median of the posterior distributions resulting from the MCMC fitting routine. Uncertainties are calculated based on the 68-th percentile highest posterior density intervals. EW and their uncertainties are computed based on the integrated flux, the continuum flux determined at the line's position and the spectroscopic redshift. In Fig.~\ref{fig:emissions} we present the best fit models.

\subsection{UV magnitude and $\beta$ slope} \label{sec:Muv}
\begin{figure*}[ht!]
\begin{minipage}{0.5\textwidth}
\centering
\includegraphics[width=\linewidth]{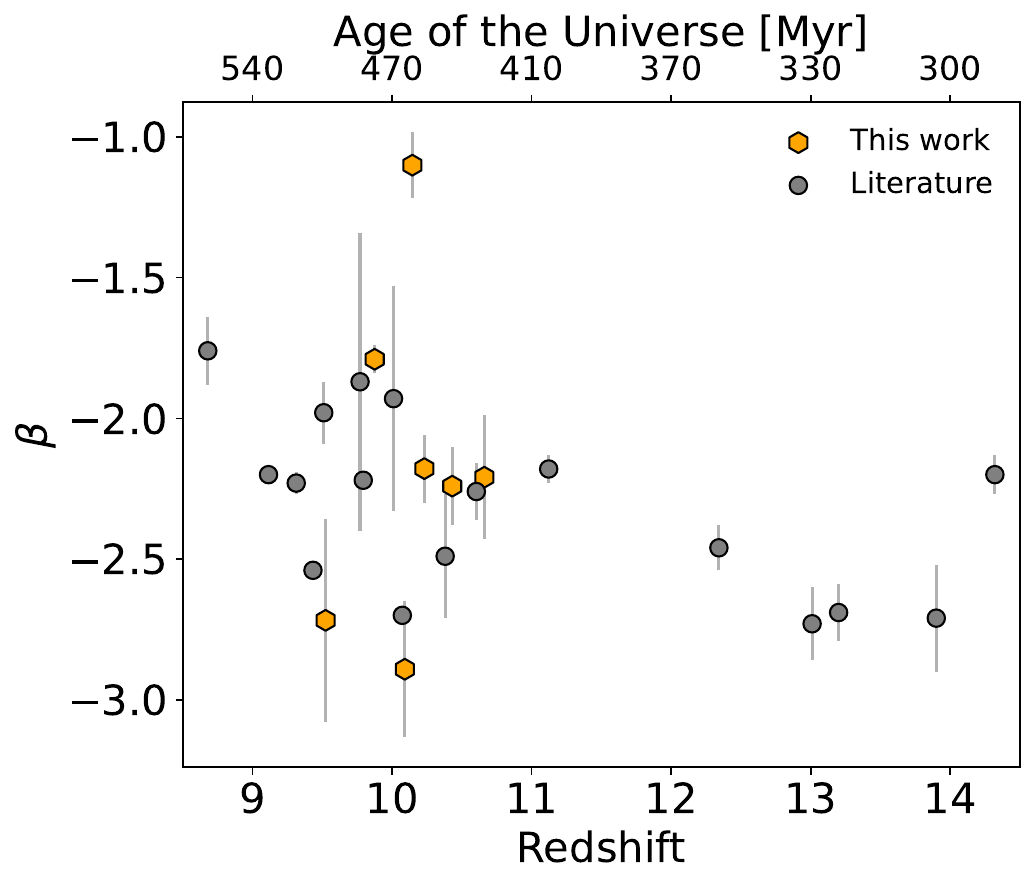}
\end{minipage}
\begin{minipage}{0.5\textwidth}
\centering
\includegraphics[width=\linewidth]{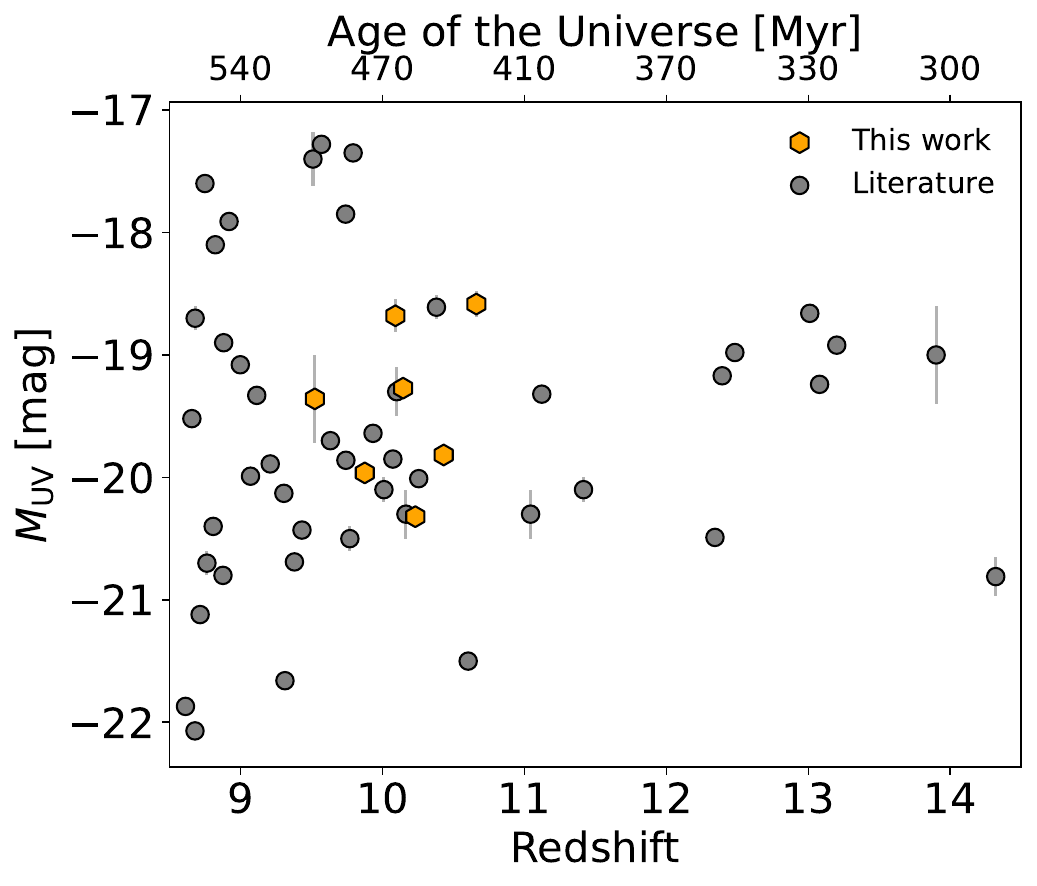}
\end{minipage}
\caption{UV $\beta$ slope and absolute magnitude $M_{\mathrm{UV}}$ evolution with redshift for z > 9 galaxies. The orange hexagons show our results, while measurements form the literature are reported as grey circles \citep{Larson2022, ArrabalHaro2023, Arrabal_Haro2023Nature, Bunker2023B, Curtis-Lake2023, D'Eugenio2023, Fujimoto2023, Goulding2023, Hsiao2023, Larson2023, Roberts-Borsani2023, Stiavelli2023, Tang2023, Wang2023, Williams2023, Boyett2024, Carniani2024, Castellano2024, Curti2024, Hainline2024, Harikane2024, Roberts-Borsani2024, Schaerer2024, Tang2024B, Witstok2024, Zavala2024}.}
\label{fig:beta&Muv}
\end{figure*}

We measure the UV $\beta$ slope by fitting a power-law model ($f_{\lambda} \propto \lambda^\beta$) to the continuum flux at 1400--2600 \AA\ rest-frame, after masking any potential emission features within the considered wavelength range. Following \cite{Heintz2024}, this is the ideal wavelength window to avoid possible damping wing contamination in the 1200--1400 \AA\ range. We employ the \textsc{emcee} package to conduct an MCMC analysis, which allowed us to identify the best-fitting parameters of the model through 100 chains and 1,000,000 steps. To constrain the UV slope, we imposed a flat prior distribution, with values ranging from -3 to 0. We derived the best-fit value of the $\beta$ slope and its uncertainty from the median and standard deviation of the posterior distribution. We compute the absolute UV magnitude, $M_{\mathrm{UV}}$ from the average continuum flux of the model in the 1450--1550 \AA\ rest-frame range. The results are reported in Table \ref{tab:summary_data}. Fig.~\ref{fig:beta&Muv} shows the $\beta$ and $M_{\mathrm{UV}}$ (corrected for magnification) versus redshift of our sample, compared to other z > 9 objects from the literature \citep{Larson2022, ArrabalHaro2023, Arrabal_Haro2023Nature, Bunker2023B, Curtis-Lake2023, D'Eugenio2023, Fujimoto2023, Goulding2023, Hsiao2023, Larson2023, Roberts-Borsani2023, Stiavelli2023, Tang2023, Wang2023, Williams2023, Boyett2024, Carniani2024, Castellano2024, Curti2024, Hainline2024, Harikane2024, Roberts-Borsani2024, Schaerer2024, Tang2024B, Witstok2024}. 

\begin{table*}[ht]
\caption{Properties of the spectroscopically-confirmed galaxies at z > 9 presented in this work.}\label{tab:summary_data}
\tiny{
\begin{tabularx}{\textwidth}{lcccccccc}
\hline \hline
\noalign{\smallskip}
\text{ID} & \text{RA} \ [\text{deg}] & \text{DEC} \ [\text{deg}] & z$_{\mathrm{spec}}$ & $\mu$ & t$^{\mathrm{exp}}$ \ [\text{s}] & M$_{\mathrm{UV}}$ \ [\text{mag}] & $\beta$ & \text{S/N > 5 emission lines}\\
\noalign{\smallskip}
\hline
\noalign{\smallskip}
\noalign{\smallskip}
GLASS-83338 & 3.454707 & -30.316890 & 9.5231 $\pm$ 0.0040 & 1.23 & 13130 & -19.36 $\pm$ 0.36 & -2.72 $\pm$ 0.36 & \parbox{1.5in}{\centering \text{[OII]}$\lambda \lambda$\text{3727,29;} \\ \text{[OIII]}$\lambda$\text{4959; [OIII]}$\lambda$\text{5007}} \\
\noalign{\smallskip}
\noalign{\smallskip}
GHZ1 & 3.511929 & -30.371859 & 9.875 $\pm$ 0.008 & 1.72 & 19701 & -19.963 $\pm$ 0.034 & -1.79 $\pm$ 0.05 & \parbox{1.5in}{\centering \text{[OII]}$\lambda \lambda$\text{3727,29;} \\ \neiii ; H$\gamma$ ; H$\beta$}\\
\noalign{\smallskip}
\noalign{\smallskip}
GLASS-z11-17225 & 3.507311 & -30.343201 & 10.09 $\pm$ 0.07 & 1.55 & 13130 & -18.68 $\pm$ 0.13 & -2.89 $\pm$ 0.24 & \parbox{1.5in}{\centering [NeIV]$\lambda$\text{2424}} \\
\noalign{\smallskip}
\noalign{\smallskip}
GHZ9 & 3.478761 & -30.345516 & 10.145 $\pm$ 0.010 & 1.36 & 19701 & -19.271 $\pm$ 0.040 & -1.10 $\pm$ 0.12 & \parbox{1.6in}{\centering NIV]$\lambda$1486; CIV$\lambda \lambda$1548,51; \\ NIII]$\lambda \lambda$1747,49; CIII]$\lambda$1909; [OII]$\lambda \lambda$3727,29; \neiii ; \\ H$\delta$ ; H$\gamma$ ; [OIII]$\lambda$4363 } \\
\noalign{\smallskip}
\noalign{\smallskip}
GHZ8 & 3.451428 & -30.321798 & 10.231 $\pm$ 0.039 & 1.23 & 13130 & -20.32 $\pm$ 0.07 & -2.18 $\pm$ 0.12 & \parbox{1.5in}{\centering [OII]$\lambda \lambda$3727,29} \\
\noalign{\smallskip}
\noalign{\smallskip}
GHZ7 & 3.451369 & -30.320717 & 10.43 $\pm$ 0.12 & 1.23 & 13130 & -19.82 $\pm$ 0.05 & -2.24 $\pm$ 0.14 & \parbox{1.5in}{\centering CIII]$\lambda$1909} \\
\noalign{\smallskip}
\noalign{\smallskip}
GHZ4$^*$ & 3.513740 & -30.351568 & 10.66 $\pm$ 0.12 & 1.74 & 32825 & -18.58 $\pm$ 0.11 & -2.21 $\pm$ 0.22 & \parbox{1.5in}{\centering NIV]$\lambda$1486 } \\
\noalign{\smallskip}
\noalign{\smallskip}
\hline
\hline
\end{tabularx}
}
\begin{tablenotes}
 \item \small $^*$The spectroscopic redshift is tentative.
 \end{tablenotes}
\end{table*}

\section{Spectroscopic confirmations at z $\sim$ 9--11} \label{sec:confirmations}
In the following, we will discuss the redshift determination of the observed galaxies and highlight the most significant features of the confirmed sources. The constraints derived from their \civ\ and \ciiired\ will be discussed in Sect.~\ref{sec:ciii}, while a more detailed investigation of other physical properties will be presented in a forthcoming paper. We report observed line fluxes, the intrinsic values can be obtained by dividing by the magnification factor $\mu$ from Table \ref{tab:summary_data}. 

\subsubsection{GLASS-83338} \label{sec:atek}
\cite{Atek2023b} identified GLASS-83338 as a photometric candidate with a redshift of z$_{\mathrm{phot}}$ = 9.55$^{+0.91}_{-0.57}$. We confirm that this galaxy is at a spectroscopic redshift of z$_{\mathrm{spec}}$ = 9.5231 $\pm$ 0.0040, based on a total effective exposure time of 13130 s. We detected the [OII]$\lambda \lambda$3727,29, the [OIII]$\lambda$4959, and the [OIII]$\lambda$5007 emission lines with a S/N of 6, 7, and 20, respectively. The $O_{32}$ ratio is 6.6 $\pm$ 1.1, which aligns well with the trend shown in Fig.~S5 of \cite{Williams2023}, who considered a low-redshift sample of extreme emission-line galaxies (EELGs). 

\subsubsection{GHZ1} \label{sec:GHZ1}
GHZ1 was first identified as a strong candidate at z$_{\mathrm{phot}}$ $\sim$ 10.5 by \cite{Castellano2022b} and \cite{Naidu2022b}, and subsequently selected by several independent analyses \citep[][]{Harikane2023, Donnan2023,Bouwens2023}. We observed this galaxy for a total of 19701 s and detected several emission lines with integrated S/N > 5: the unresolved [OII]$\lambda \lambda$3727,29 doublet, \neiii , H$\gamma$, and \hb. The spectroscopic redshift is unambiguously constrained at z = 9.875 $\pm$ 0.008. The galaxy also shows a clear \lya-break and significant NIII]$\lambda \lambda$1747,49 and the unresolved [NeIII]$\lambda$3967 and H$\epsilon$ emission lines.\\
This galaxy was also observed by the Atacama Large Millimeter/submillimeter Array (ALMA). The spectroscopic solution obtained from the NIRSpec data is in perfect agreement with the lack of a clear detection of the [OIII]88$\mu$m line in the redshift range 10.10 < z < 11.14 from the ALMA data \citep{Yoon2023}. 

\subsubsection{GLASS-z11-17225} \label{sec:mcleod}
GLASS-z11-17225 was identified as a galaxy candidate at z$_{\mathrm{phot}}$ = 10.7$^{+0.3}_{-0.6}$ by \cite{McLeod2024}. Our 13130 s spectroscopic follow-up confirms it as a galaxy at z$_{\mathrm{spec}}$ = 10.09 $\pm$ 0.07. This redshift determination is based on the detection of the unresolved [NeIV]$\lambda \lambda$2422,24 doublet with S/N of 5.9, while the significant [OII]$\lambda \lambda$3727,29 has S/N = 4.6. The [NeIV]$\lambda \lambda$2422,24 feature is considered a strong AGN indicator \citep[e.g.,][]{Feltre2016, LeFevre2019}, requiring 63.5 eV from the ionizing source to be produced. This line was identified by \cite{Maiolino2024} in GN-z11 and discussed as clear evidence of the AGN nature of that source, in the context of several high-ionization lines detections. However, in our case, the absence of other significant emission lines prevents us from conclusively proving the AGN nature of GLASS-z11-17225.\\
Given the lack of other significant emission lines aside from the [NeIV]$\lambda \lambda$2422,24 and the [OII]$\lambda \lambda$3727,29, we also investigated the possibility that this source could be a low-redshift interloper. The only other viable redshift solution is z$_{\mathrm{spec}}$ $\sim$ 2.75, inferred from the Balmer Break and a possible absorption feature at the expected position of the unresolved CaK and CaH lines. However, in this low-redshift scenario, the previously discussed emission features would not correspond to any known lines and no relevant emission lines would be detected, making this redshift solution less likely. The absence of Paschen, Balmer, and the \oiiidoub\ lines would imply this is a passive galaxy, similar to the one presented by \cite{Carnall2023}.\\
Moreover, no significant emission is detected blueward of the break. The constraint on the F115W flux implies a 1$\sigma$ upper limit of $\sim$31 mag, corresponding to a $>$2.4 magnitude break. Consistently, the photometric redshift solutions \citep[e.g.,][Merlin et al., subm]{McLeod2024} strongly point to the high-redshift scenario. Therefore, hereafter we adopt the z$_{\mathrm{spec}}$ = 10.09 $\pm$ 0.07, which is compatible with both the \lya-break and the [NeIV]$\lambda \lambda$2422,24 and [OII]$\lambda \lambda$3727,29 doublets.

\subsubsection{GHZ9} \label{sec:GHZ9}
C23 identified GHZ9 as a photometric candidate with a redshift of z$_{\mathrm{phot}}$ = 9.35$^{+0.77}_{-0.35}$. We confirm that this galaxy has a spectroscopic redshift of z$_{\mathrm{spec}}$ = 10.145 $\pm$ 0.010, based on a total effective exposure time of 19701 s. This redshift solution is based on several emission lines with S/N > 5:  NIV]$\lambda$1486, unresolved CIV$\lambda \lambda$1548,51, and NIII]$\lambda \lambda$1747,49 doublets, CIII]$\lambda$1909, unresolved [OII]$\lambda \lambda$3727,29 doublet, \neiii, H$\delta$, H$\gamma$ and [OIII]$\lambda$4363. The \neiii\ line has a S/N of 10, while CIII]$\lambda$1909 and CIV$\lambda \lambda$1548,51 each have S/N values $\sim$ 9. 
For the CIII]$\lambda$1909 line, we find an integrated flux of (11.0 $\pm$ 1.2) $\times 10^{-19}$ erg/cm$^2$/s and a rest-frame equivalent width of (48 $\pm$ 5) \AA\ , while for CIV$\lambda \lambda$1548,51 we obtain (17.3 $\pm$ 1.9) $\times 10^{-19}$ erg/cm$^2$/s and (65 $\pm$ 7) \AA\ . We also identify a significant OIII]$\lambda \lambda$1661,66 doublet with a S/N = 3.5. We note that the redshift solution is compatible with the \lya-break feature. \\
Based on 2.1 Ms deep Chandra observations, \citet{Kovacs2024} detected 0.5–3 keV emission at the position of GHZ9, which they interpret as evidence of this source being an AGN with a super-massive black hole (SMBH) of 8.0$^{+3.7}_{-3.2} \times$10$^7$M$_{\odot}$. We will discuss this scenario in the light of CIII]$\lambda$1909, and CIV$\lambda \lambda$1548,51 EWs and line ratios in Sect.~\ref{sec:ciii}. 


\subsubsection{GHZ8} \label{sec:GHZ8}
GHZ8, another photometric candidate from C23 at z$_{\mathrm{phot}}$ = 10.85$^{+0.45}_{-0.57}$, is confirmed to have a spectroscopic redshift of 10.231 $\pm$ 0.039, leveraging a total exposure time of 13130 s. This source is the brightest in our sample, with $M_{\mathrm{UV}}$ = (-20.32 $\pm$ 0.07) mag. We identify a clear \lya-break, the unresolved [OII]$\lambda \lambda$3727,29 doublet, the blended H$\epsilon$ with [NeIII]$\lambda$3967 emissions and H$\gamma$, with S/N of 5.0, 3.7, and 3.2, respectively. The low signal to noise ratio of the combined H$\epsilon$ and [NeIII]$\lambda$3967 lines prevents deblending the two emission features. \\

\subsubsection{GHZ7} \label{sec:GHZ7}
The photometric solution for GHZ7 identified by C23 was z$_{\mathrm{phot}}$ = 10.62$^{+0.55}_{-1.02}$. We observed this galaxy for a total of 13130 s and find its spectroscopic redshift is z$_{\mathrm{spec}}$ = 10.43 $\pm$ 0.12. This redshift solution is based on the CIII]$\lambda$1909 emission with S/N > 5 and supported by the clear \lya-break feature, while CIV$\lambda \lambda$1548,51, [NIV]$\lambda$1718, and [NeV]$\lambda$3346 are significant emission lines with S/N > 3. In particular, the CIII]$\lambda$1909 and CIV$\lambda \lambda$1548,51 lines have S/N = 5.9 and 3.2, with observed fluxes of (7.7 $\pm$ 1.3) $\times 10^{-19}$ erg/cm$^2$/s and (5.8 $\pm$ 1.8) $\times 10^{-19}$ erg/cm$^2$/s, respectively. The rest-frame equivalent widths of CIII] is (33 $\pm$ 6) \AA, while for the CIV we obtain (17 $\pm$ 5) \AA. \\
Interestingly, we detect [NeV]$\lambda$3346 at S/N = 3 and tentative [NeV]$\lambda$3426 at S/N $\sim$ 2. The high ionization potential of [NeV] ($\sim$ 97.12 eV), makes this feature unlikely to be produced by stellar populations \citep[e.g.,][]{Mignoli2013, Feltre2016}. Recently, \cite{Curti2024} and \cite{Chisholm2024} reported [NeV] detections at $\geq 3\sigma$ in sources at z = 9.43 and z = 5.6, respectively and discussed the AGN origin of this line. 

\subsubsection{GHZ4} \label{sec:GHZ4}
GHZ4 was first identified as a high-redshift candidate by \citet{Castellano2022b} and \citet{Harikane2023}. C23 estimated a photometric redshift of z$_{\mathrm{phot}}$ = 10.27$^{+1.2}_{-1.4}$ for this object. From our observations and a total exposure time of 32825 s we confirm this is a galaxy at z$_{\mathrm{spec}}$ = 10.66 $\pm$ 0.12 based on the NIV]$\lambda$1486 emission line with S/N = 5. The CIII]$\lambda$1909, and H$\gamma$ are significant features with S/N of 4, and 3.6, respectively. Among these detected lines, CIII]$\lambda$1909 has an observed flux of (2.8 $\pm$ 0.7) $\times 10^{-19}$ erg/cm$^2$/s with a EW$_0$ = (26 $\pm$ 7) \AA . The spectrum exhibit a \lya-break compatible with the same redshift solution, with no significant flux detected at shorter wavelengths. We considered the possibility that this source might be a low-redshift interloper. The only other plausible redshift solution is z$_{\mathrm{spec}}$ $\sim$ 2.86, suggested by the Balmer Break. However, in this low-redshift scenario, the emission features previously discussed would not correspond to any recognized lines, and no significant emission lines would be detected, making this redshift solution less likely. In our sample, this is the only case for which the spectrum shows low S/N continuum. 
Hereafter, we will consider the redshift solution z$_{\mathrm{spec}}$ = 10.66 $\pm$ 0.12 of GHZ4 as a tentative one. 

\subsection{Other high-redshift targets} \label{sec:interlopers}

\begin{figure}[t]
\centering
\includegraphics[width=\linewidth]{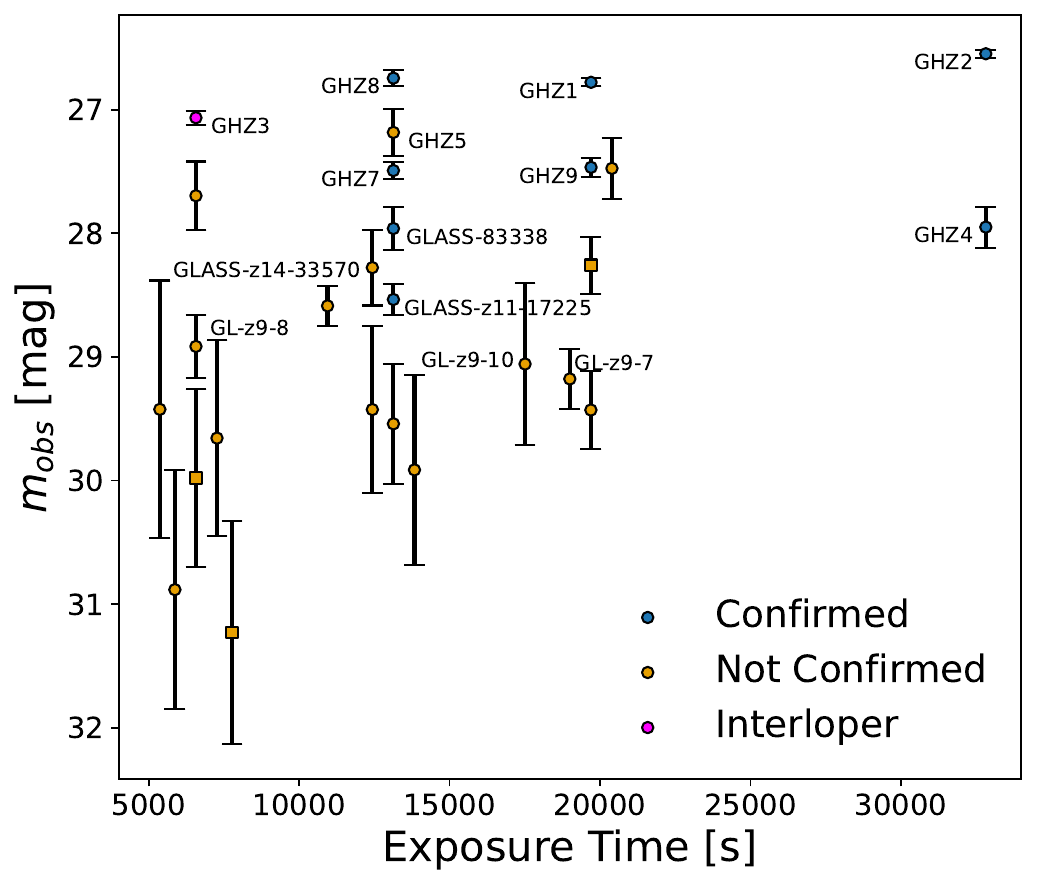}
\caption{Observed magnitudes from NIRCam photometry and total exposure time from NIRSpec PRISM data are shown for all the 27 photometric candidates considered in this study. We report the F200W measurement (circle) whenever available, while the F277W band (square) is shown in all the other cases. Data are color-coded based on the spectroscopic redshift result: confirmed (blue), not confirmed (orange), and interloper (magenta). Some of the points have been slightly shifted along the x-axis for an easier visualization.} \label{fig:candidates}
\end{figure} 

We targeted other six objects previously presented as reliable high-redshift candidates on the basis of NIRCam photometry: GHZ3 and GHZ5 from \citet{Castellano2022b}, GL-z9-7, GL-z9-8 and GL-z9-10 \citep{Harikane2023}, and GLASS-z14-33570 \citep{McLeod2024}.\\
%
%
The spectrum of GHZ3 shows a break at 10070 \AA\ and no emission lines. If we consider this feature a Balmer (\lya)-break, this galaxy has a spectroscopic solution of z $\sim$ 1.76 (z $\sim$ 7.28) (see Fig.~\ref{fig:GHZ3_new}). We note that the z $\sim$ 1.76 solution is compatible with the photometric redshift probability distribution P(z) discussed in \citet{Castellano2022b}. In fact, at variance with our main sample from C23, GHZ3 and GHZ5 were found to have a non-negligible probability at $z\lesssim$ 3. \\
We do not find a clear redshift solution for the other candidates mentioned above, due to the low S/N spectra.
Similarly, we do not identify spectroscopic features for any of the other 13 observed sources with photometric redshift z $\gtrsim$ 9 from \cite{Paris2023} or Merlin et al. (subm). In Fig.~\ref{fig:candidates} we show the observed magnitudes as a function of the exposure time for all the photometric candidates and spectroscopic confirmations from our study. We note that our main targets from C23 have mag $\lesssim$ 28 in the F200W band, and all the above mentioned confirmed objects have mag $\lesssim$ 28.5, while the objects without a clear redshift solution are generally fainter and have poorer constraints on their photometric redshifts.

In summary, all the six strong high-redshift candidates identified in \cite{Castellano2022b} and C23 (i.e., GHZ1, GHZ2, GHZ4, GHZ7, GHZ8, and GHZ9) have been confirmed by our spectroscopic observations. The confirmation success rate of our spectroscopic follow-up of bright z$\sim$10 candidates is 75\% when including also the objects with a double photometric redshift solution (i.e., GHZ3 and GHZ5) from \cite{Castellano2022b}, and conservatively considering the latter as an interloper. 

\subsection{Constraints on AGN contribution from carbon lines}\label{sec:ciii} 
\begin{figure*}[ht!]
\begin{minipage}{0.5\textwidth}
\centering
\includegraphics[width=\linewidth]{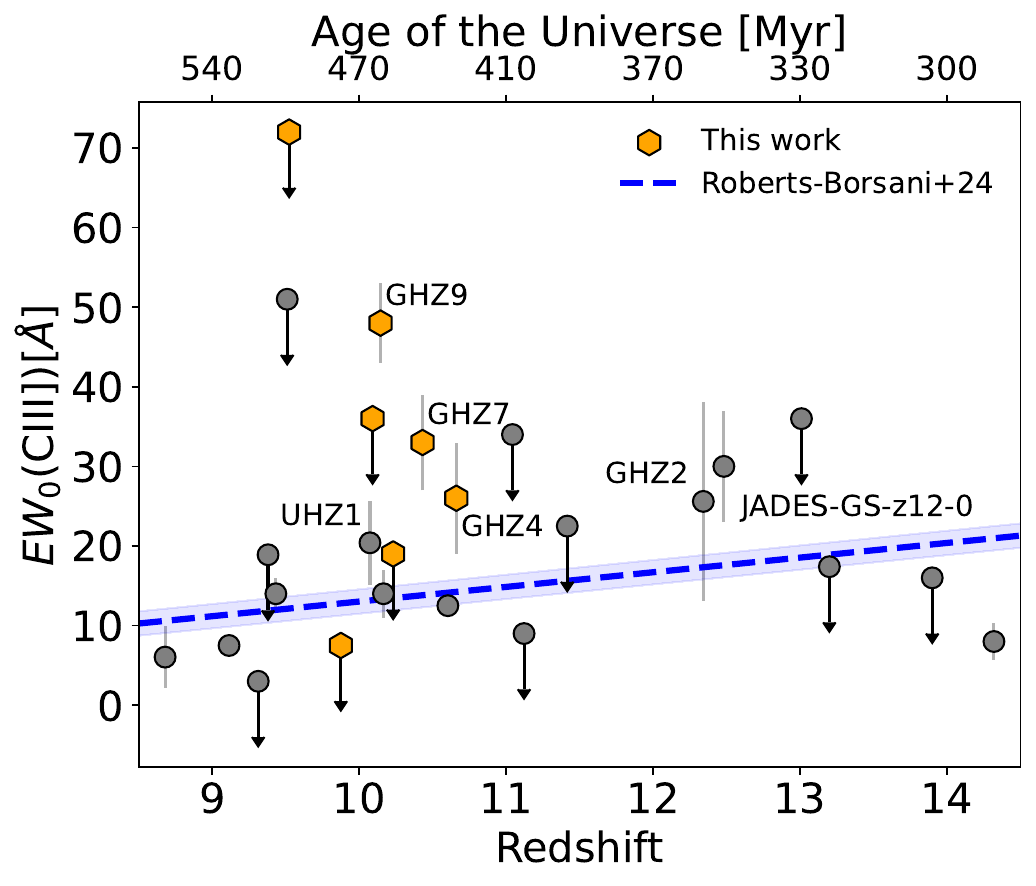}
\end{minipage}
\begin{minipage}{0.5\textwidth}
\centering
\includegraphics[width=\linewidth]{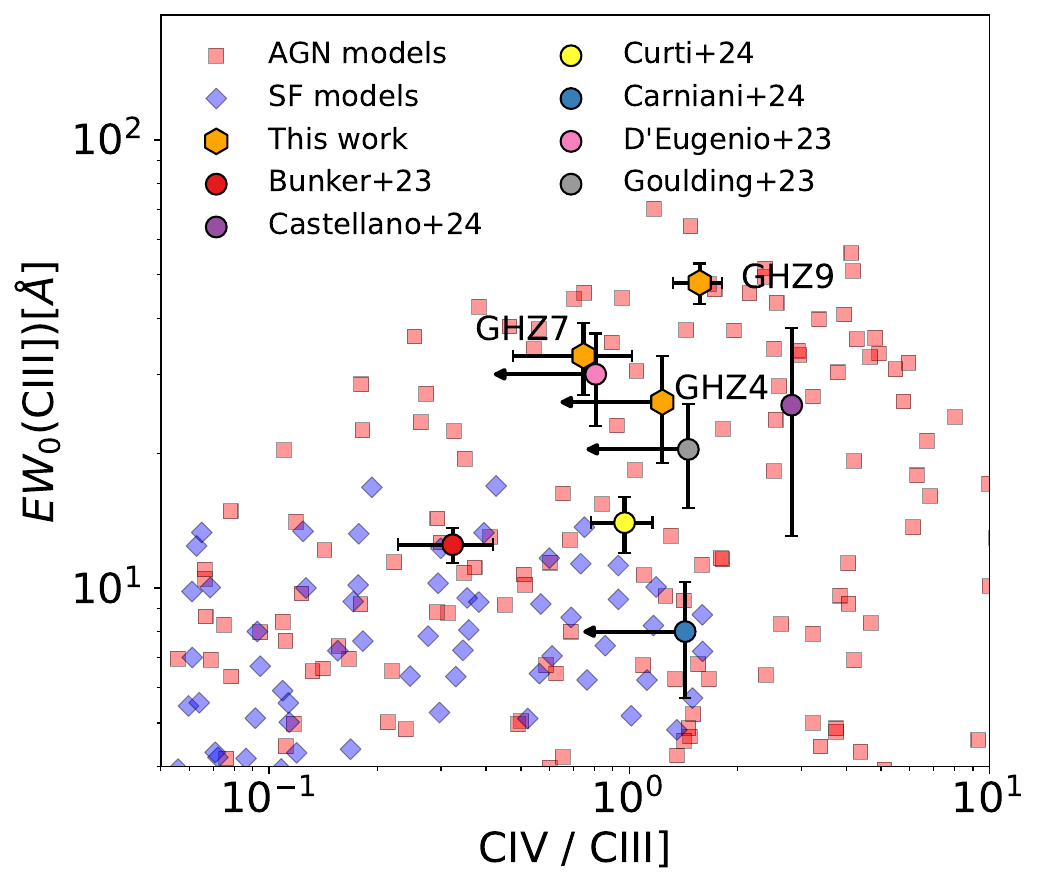}
\end{minipage}
\caption{Left: Redshift evolution of the EW$_0$ of CIII]$\lambda$1909. The best-fit relation found by \cite{Roberts-Borsani2024} is indicated by the blue dashed line. We present both measurements and upper limits. The orange hexagons represent our results, while results from the literature are shown as grey circles \citep{ArrabalHaro2023, Bunker2023B, D'Eugenio2023, Hsiao2023, Goulding2023, Larson2023, Stiavelli2023, Williams2023, Boyett2024, Carniani2024, Castellano2024, Curti2024, Hainline2024, Schaerer2024, Witstok2024}. Right: Diagnostic diagram based on CIII]$\lambda$1909 and CIV$\lambda \lambda$1548,51 emission lines. The AGN and star-forming models from \cite{Nakajima2022} are displayed as red squares and blue diamonds, respectively. We include a subset of the sources shown in the left plot, where measurements or limits for CIV$\lambda \lambda$1548,51 are available.}
\label{fig:CIII}
\end{figure*}

Three galaxies in our sample, namely GHZ4, GHZ7, and GHZ9 show prominent CIII]$\lambda$1909 emission. While such intense emission is quite rare among star-forming galaxies at low to intermediate redshifts, appearing in only 1\% of the cases at 2 < z < 4, \citep[e.g.,][]{LeFevre2019}, at z > 6 high EW values of CIII]$\lambda$1909 are sometimes observed \citep[e.g.,][]{Mainali2017, Stark2017, Hutchison2019}, leading to a puzzle regarding the nature of these sources. High CIII]$\lambda$1909 emission might be due to either an AGN powering mechanism or low gas metallicity \citep[e.g.,][]{Jaskot2016, Nakajima2018a, Llerena2022}. \\
Recently, \cite{Roberts-Borsani2024} found a trend of average CIII]$\lambda$1909 EW increasing with redshift, utilizing spectral stacks of many hundreds of spectroscopically-confirmed sources, and discussed this in the context of decreasing metallicity. In the left panel of Fig.~\ref{fig:CIII} we compare these results to the EW values of CIII]$\lambda$1909 for our sample, as well as a few additional high redshift galaxies from recent literature. Indeed GHZ4, GHZ7, and GHZ9 exhibit EW values much above the average trend. \\ 
To determine the nature of the sources powering the carbon emissions in our sample, we employ the CIV/CIII] versus EW(CIII]) diagnostic diagram and compare our results with the AGN and star-forming (SF) models by \cite{Nakajima2022}. In the right panel of  Fig.~\ref{fig:CIII} we can see that GHZ4, GHZ7, and GHZ9 are consistent only with AGN models, similar to GHZ2 \citep{Castellano2024}, JADES-GS-z12-0 \citep{D'Eugenio2023}, and UHZ1 \citep{Goulding2023, Bogdan2024}. However, it is important to note that, for example in the case of GHZ2, other diagnostics diagrams gave discrepant results on the nature of this source \citep{Castellano2024}. 
In contrast, GS-z9-0 \citep{Curti2024}, GN-z11 \citep{Bunker2023B, Maiolino2024}, and JADES-GS-z14-0 \citep{Carniani2024} occupy the region of the diagram where both AGN and SF models could explain their emissions. We note that the AGN nature of GHZ7 is also suggested by the presence of the [NeV]$\lambda$3346 emission. However, the most striking case is that of GHZ9, which was also associated with a Chandra X-ray detection by \cite{Kovacs2024}, further corroborating the AGN nature of this source. A more detailed analysis of this object will be presented in a forthcoming paper (Napolitano et al., in prep.).\\

\section{Confirmation of a high abundance of z$\sim$10 galaxies} \label{sec:abundance}

\begin{figure}[t]
\centering
\includegraphics[width=\linewidth]{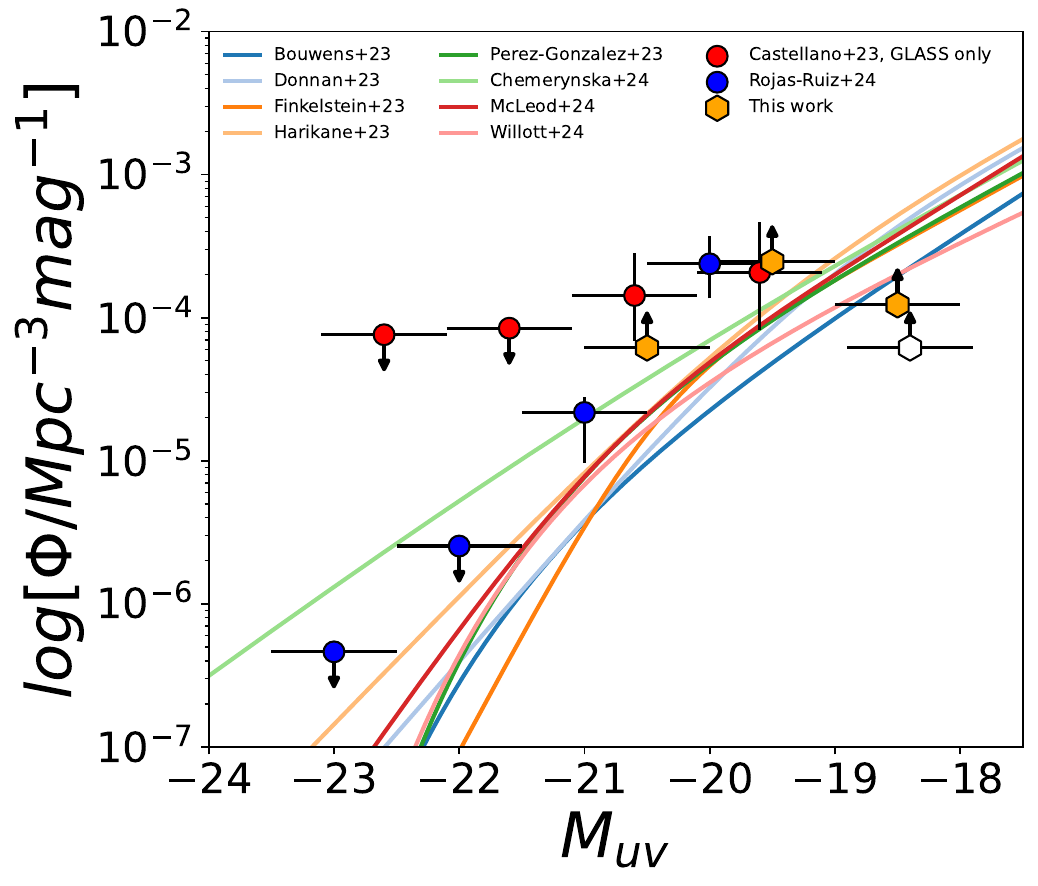}
\caption{The UV luminosity function at z $\sim$ 10 in the GLASS-JWST ERS field. The orange hexagons show the binned UV LF obtained from the spectroscopic sample discussed in this work. The open hexagon shows the estimate at -19 $\leq M_{\mathrm{UV}} \leq$ -18 after removing GHZ4, whose redshift is not secure. Spectroscopic estimates are represented as lower limits to highlight that they are not corrected for the incompleteness of the photometric selection. Red circles and error-bars indicate the previous photometric estimate by C23. Blue points show the binned UV LF at z = 8.4--10 from the BoRG-JWST survey \citep[][]{Rojas-Ruiz2024}. Data points have been slightly shifted in $M_{\mathrm{UV}}$ for an easier visualization. The solid coloured lines show estimates of the z$\sim$10 UV LF from the literature:
\cite{Bouwens2023}, \cite{Donnan2023}, \cite{Finkelstein2023b}, \cite{Harikane2023}, \cite{PerezGonzalez2023}, \cite{Chemerynska2024}, \cite{McLeod2024}, and \cite{Willott2024}. } \label{fig:UV_LF}
\end{figure}

We exploit the successful confirmation of seven galaxies at 9 < z < 11 to constrain the z$\sim$10 UV luminosity function in the field. 
As discussed in Sect.~\ref{sec:Nircam}, these galaxies were selected as high-redshift sources from the two NIRCam parallel fields of the GLASS-JWST ERS program. Although the NIRSpec pointings do not cover the entire region, their positioning was chosen as to include all high priority candidates from C23. We thus conservatively consider here the area of the entire GLASS-JWST ERS NIRCam field. After correcting for lensing on the basis of the model by \cite{Bergamini2023} as described in C23, we estimate that the field spans a comoving volume of 16,250 cubic comoving Mpc (cMpc$^3$) at 9.5 $\leq$ z $\leq$ 11. We use the demagnified rest-frame $M_{\mathrm{UV}}$ 
to estimate the UV LF in two magnitude bins at -21 $\leq M_{\mathrm{UV}} \leq$ -19. The results are shown in Fig.~\ref{fig:UV_LF}. Our spectroscopic estimates are represented as lower limits to highlight that they are not corrected for the incompleteness of the photometric selection. The comparison with recent estimates of the UV luminosity function in the same redshift range \citep[e.g.,][]{Bouwens2023, Donnan2023, Finkelstein2023b, Harikane2023, PerezGonzalez2023, Chemerynska2024, McLeod2024, Willott2024} confirms that the number density of z $\sim$ 10 galaxies in the GLASS-JWST ERS field is a factor of $>$ 3 higher than the average, as reported by C23 on the basis of the photometric sample. While the limit on the number density at -19 < $M_{\mathrm{UV}}$ < -18 is formally consistent with values from the literature, this magnitude range is clearly affected by a high incompleteness in both the photometric selection and in the sampling of the spectroscopic targets, given that the NIRSpec observations were designed to follow-up the brighter sample of LBGs from C23.



\subsection{Is the high number density due to a proto-structure at z$\sim$10?} \label{sec:protocluster}

The unique high abundance of galaxies at z $\sim$ 10 detected in the GLASS field may be attributed to the presence of a structure of galaxies in the process of formation. Here, we further constrain this scenario by investigating the nature of this densely populated galaxy assembly.\\
To assess whether two galaxies belong to the same proto-cluster, we considered the distance criterion defined by \cite{Chiang2017}, which considers the boundary of the structure to be the average distance $\langle R_L \rangle$ at which the membership probability drops to 50\%. Briefly, they selected a mass-complete sample of galaxy clusters at z = 0 with virial M$_{200} >$10$^{14}$M$_{\odot}$ in the semi-analytic models by \citet{Henriques2015} and \citet{Guo2013}, and tracked the evolution of their progenitors back in cosmic time to constrain the size of parent proto-clusters at high-redshift. They found that $\langle R_L \rangle$ reaches an asymptotic value of $\sim$ 10.3 cMpc at redshift higher than 6--7. The corresponding physical radius of a proto-cluster at z = 10 would be 0.93 pMpc.\\
We independently derived the $\langle R_L \rangle$ values using the TNG300 simulation \citep[see][]{Marinacci2018, Naiman2018, Nelson2018, Pillepich2018, Springel2018} to better constrain the dependence of this distance on the virial halo mass in the range 10$^{14}$M$_{\odot}$ < M$_{200}$ < 10$^{15}$M$_{\odot}$. We selected all the galaxies within ten virial radii (R$_{200}$) from the cluster center at z = 0, and tracked the positions of the progenitors of both cluster members and field galaxies back to z = 10, following the evolution of $\sim$ 3500 proto-clusters. Our $\langle R_L \rangle$ values are in good agreement with the result reported by \cite{Chiang2017} for M$_{200}$ > 10$^{14.5}$M$_{\odot}$, while for lower masses we obtain slightly smaller distances (0.4--0.6 pMpc). We also found no significant evolution of $\langle R_L \rangle$ in the redshift range 7 < z < 10. 

\begin{figure}[t]
\centering
\includegraphics[width=\linewidth]{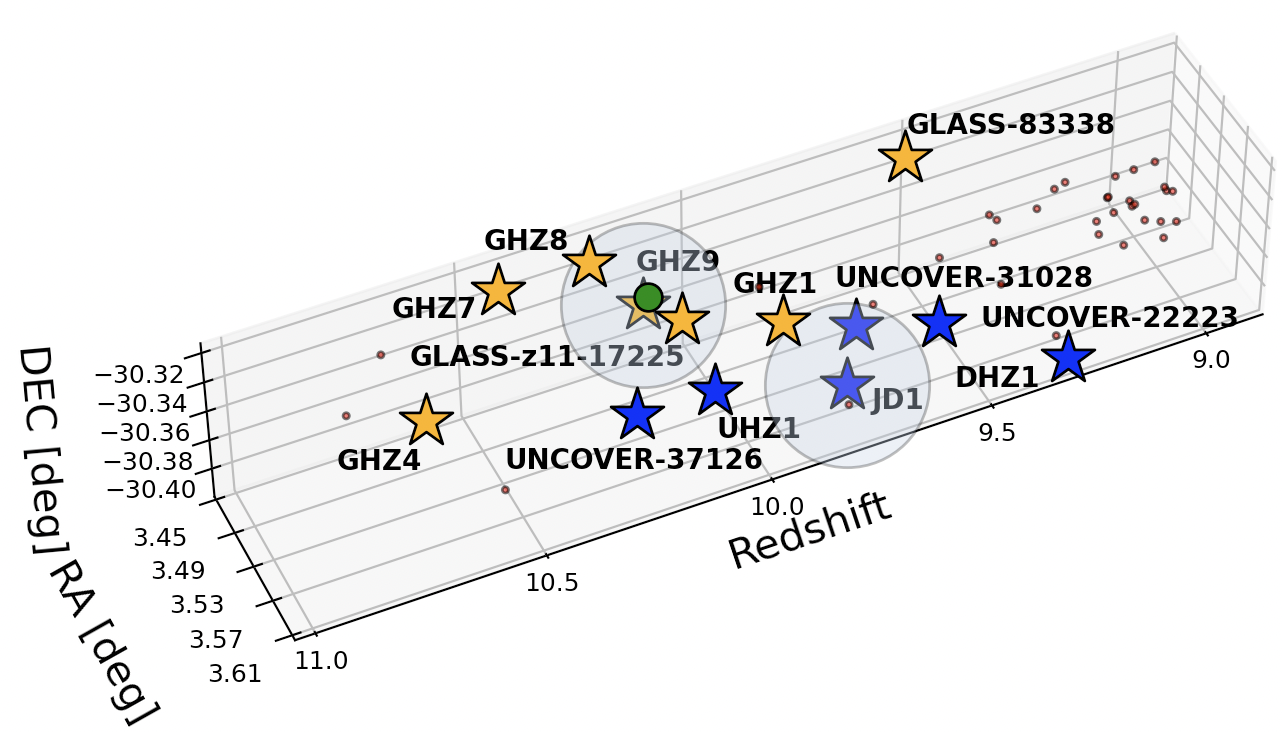}
\caption{The position of the z $\sim$ 10 spectroscopically-confirmed galaxies in the Abell-2744 field in redshift and sky coordinates space. The orange stars show the position of the galaxies identified in this work, with the green central point representing the barycenter of their 3-d distribution. The blue stars represent the other spectroscopically-confirmed objects from literature \citep[][]{Fujimoto2023, Goulding2023, Roberts-Borsani2023, Boyett2024}. We highlight the possible proto-clusters found in the proximity of GHZ9 and JD1. The red circles show the position of additional m(F200W) < 28.5 candidates with photometric redshift 9 $\leq$ z $\leq$ 11 in the parent catalog by Merlin et al. (subm). We note that the distribution on the sky of these additional candidates is due to the configuration of the NIRCam pointings (see Fig.~\ref{fig:pointing}).} \label{fig:3DStructure}
\end{figure}

To compare with observations, we first assessed the scenario in which the seven spectroscopically-confirmed galaxies are part of a unique structure. To this aim, we computed the distance between each of the objects and the barycenter of the sample following the distance solution by \cite{Liske2000}.
We use our spectroscopic redshift measurements and sky coordinates projected onto the source plane on the basis of the lensing model by \citet{Bergamini2023}. The position of our targets and of the relevant barycenter in redshift and sky coordinates space is shown in Fig.~\ref{fig:3DStructure}. We found that only GHZ9 and GLASS-z11-17225 would meet the criterion for proto-cluster membership, with GHZ8 located at twice the expected physical distance $\langle R_L \rangle$ defined by \cite{Chiang2017}. 
All other galaxies are positioned 5-10 times farther from the barycenter than the radius $\langle R_L \rangle$, indicating that our sample does not populate a single progenitor of a present-day cluster. 
We repeated this exercise considering all the spectroscopic z$\sim$10 galaxies in the Abell-2744 field, i.e. UNCOVER-31028, UNCOVER-22223, UNCOVER-37126 \citep{Fujimoto2023}, UHZ1 \citep{Goulding2023}, Gz9p3/DHZ1 \citep{Boyett2024}, and JD1 \citep{Roberts-Borsani2023}, and the conclusion does not change.

We then computed the physical distances between each pair of galaxies in our sample to explore the possibility that it includes both proto-clusters that are physically associated and "field" galaxies that are not part of these emerging structures.
With this method, we find that
GHZ9 and JD1 each have two companion galaxies. GHZ9 is paired with GLASS-z11-17225 and GHZ8, as already identified, while JD1 is associated with UNCOVER-31028, and GHZ1. We note that a less stringent estimate of the proto-cluster size, defined as the distance at which the membership probability drops to 16\% in the TNG300 simulation, would include GHZ9, GLASS-z11-17225, GHZ8, and UHZ1 as part of a single progenitor of a structure with M$_{200}$ > 10$^{14.6}$M$_{\odot}$ at z=0. The same applies to the galaxy assembly composed of JD1, UNCOVER-31028, and GHZ1.\\

In summary, our analysis indicates that the confirmed galaxies in the GLASS-JWST ERS field are scattered through too large a volume to be likely part of a single proto-cluster. However, the presence of two more clustered sub-samples may indicate that the Abell-2744 field hosts z $\sim$ 10 structures in their early stages of formation. In this respect, the enhanced clustering around the AGN candidate GHZ9 is suggestive of the overdensities typically found around active galaxies at high redshift \citep[e.g.,][]{Mignoli2020, Wang2024}. \\
A similar analysis of the earliest formation stages of more extended structures, such as super-clusters and \textquote{walls} \citep[e.g.,][]{Galametz2018, Cucciati2018}, is needed to determine whether these objects are part of a larger, forming superstructure along with the remaining GLASS galaxies, or if the latter should be considered simple \textquote{field} galaxies. Finally, as shown in Fig.~\ref{fig:3DStructure}, the Abell-2744 region hosts additional 34 bright (m(F200W) < 28.5) sources with photometric redshift in the z $\simeq$ 9--11 range (Merlin et al., subm). 
Future NIRSpec observations targeting these candidates could yield crucial information on the possible presence of a proto-cluster, or of a proto-supercluster at z $\sim$ 10 in the GLASS field.

\section{Summary and conclusions} \label{sec:Summary}
We have presented the results of a systematic follow-up of z > 9 candidate galaxies in the GLASS-JWST ERS field based on the JWST NIRSpec program GO-3073. The sample included all LBG candidates from \citet{Castellano2023} and additional sources from the literature or selected on the basis of their photometric redshift, for a total of 27 candidates. In this work, we spectroscopically confirm six sources (GHZ1, GHZ7, GHZ8, GHZ9, GLASS-83338, and GLASS-z11-17225) with secure redshifts at z = 9.52--10.43, while GHZ2 at z = 12.34 was presented in \cite{Castellano2024}. One additional object (GHZ4) does not show a convincingly clear set of emission features with S/N > 3, but is estimated to be most likely at z $\simeq$ 10.66 on the basis of the \lya-break feature and a single emission line. We also find GHZ3 is a low-redshift interloper, based on the break feature in its spectrum, marginally compatible with the photometric redshift probability distribution discussed in \citet{Castellano2022b}. The overall success rate of our follow-up on bright z$\sim$10 candidates from C23 and \citet{Castellano2022b} is 75\%.\\
The spectroscopically-confirmed sample indicates that the number density of galaxies with rest-frame demagnified -21 $\leq M_{\mathrm{UV}} \leq$ -19 at z $\simeq$ 9--11 in this field is $\gtrsim$3 times higher than estimates obtained in other JWST surveys. This is in agreement with the results obtained by \citet{Castellano2023} and \citet{McLeod2024} from photometric-selected samples. 
Regarding the excess in the GLASS region, we find that the positions of the objects in redshift and angular space are not consistent with them all being part of a single progenitor of present-day galaxy clusters. 
By considering all the spectroscopic z$\sim$10 sources in the Abell-2744 field, we identify two potential galaxy proto-clusters centered around GHZ9 and JD1, with relative separations between their members of $\simeq$1-2 pMpc. GHZ9 and JD1 have two companions each: GHZ9 is paired with GLASS-z11-17225 and GHZ8, while JD1 is associated with UNCOVER-31028, and GHZ1.
We speculate that the high number of objects in the field could be explained either by the chance superposition of a proto-cluster and field galaxies or by clustering on large scales, representing the initial formation phase of giant structures observed at lower redshifts \citep[e.g.,][]{Cucciati2018}.
Most importantly, the GLASS spectroscopic sample provides an interesting clue on the physical mechanisms behind the \textquote{excess} of galaxies at cosmic dawn measured by several independent JWST surveys \citep[e.g.,][]{Harikane2023, Finkelstein2023b, Donnan2023}. In fact, our sample includes three objects with EW(\ciii)>20\AA\ that occupy a region compatible with AGN emission in the EW(CIII]) vs CIV/CIII] diagram. One of these objects, GHZ9 at z = 10.145, has a red UV slope $\beta=$-1.10 $\pm$0.12 and it has been previously associated to a point-like X-ray source in deep Chandra data by \citet{Kovacs2024}, further supporting its AGN nature. In this context, the slow evolution of the UV LF beyond z $\simeq$ 9 may be attributed to an increased contribution of accretion onto SMBH to the UV emission of the sources \citep[e.g.,][]{Harikane2023b, Maiolino2023b}. These sources may be significant contributors to the earliest stages of cosmic reionization \citep[][]{Giallongo2015, Dayal2024, Madau2024}. It will be fundamental to assess this scenario using higher-resolution spectroscopy to detect broad emission features or faint, highly-ionized emission lines.

\begin{acknowledgements}
This work is based on observations made with the NASA/ESA/CSA {\it James Webb Space Telescope (JWST)}. The JWST data presented in this article were obtained from the Mikulski Archive for Space Telescopes (MAST) at the Space Telescope Science Institute. The specific observations analyzed are associated with program JWST-GO-3073 and can be accessed via \href{https://doi.org/10.17909/4r6b-bx96}{DOI}.  We thank Tony Roman (Program Coordinator) and Glenn Wahlgren (NIRSpec reviewer) for the assistance in the preparation of GO-3073 observations. We acknowledge support from INAF Mini-grant "Reionization and Fundamental Cosmology with High-Redshift Galaxies",  from the INAF Large Grant 2022 “Extragalactic Surveys with JWST” (PI Pentericci), from the INAF Mini Grant 2022 “Tracing filaments through cosmic time” (PI Vulcani), from PRIN 2022 MUR project 2022CB3PJ3 - First Light And Galaxy aSsembly (FLAGS) funded by the European Union – Next Generation EU and from the European Union – NextGenerationEU RFF M4C2 1.1 PRIN 2022 project 2022ZSL4BL INSIGHT. Support was also provided by NASA through grant JWST-ERS-1342. TT and SRR acknowledge support from NASA through grant JWST-GO-3073.

\end{acknowledgements}

\bibliographystyle{aa}
\bibliography{biblio.bib}

\appendix
\section{}\label{sec:app_figures}
\begin{figure*}[ht!]
    \centering
    \includegraphics[width=0.75\textwidth]{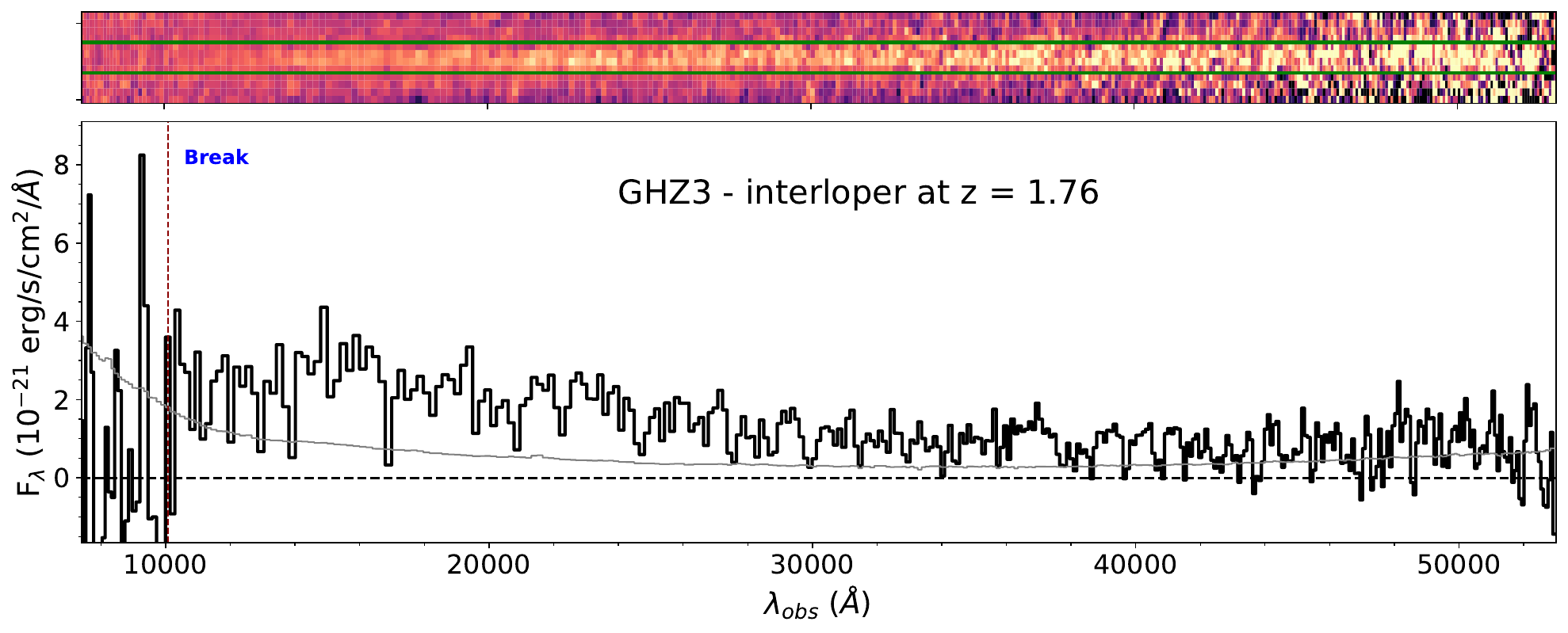}
    \caption{Observed 2D (upper panels) and 1D (lower panels) PRISM spectra of GHZ3. Symbols are the same as in Fig.~\ref{fig:spectra}. The Balmer (Lyman)-break feature is highlighted by the dashed vertical line, corresponding to z $\sim$ 1.76 ($\sim$ 7.28). The lower redshift solution is compatible with the photometric redshift probability distribution P(z) discussed in \cite{Castellano2022b}.}
    \label{fig:GHZ3_new}
\end{figure*}

In Fig.~\ref{fig:emissions} we present all the fitted profiles of significant lines. The galaxies are listed in order of increasing redshift, and the lines are organized by ascending wavelength.\\
In Fig.~\ref{fig:GHZ3_new} we show the spectrum of GHZ3. The Balmer (\lya)-Break feature at 10070 \AA\ implies a spectroscopic solution of z $\sim$ 1.76 (z $\sim$ 7.28). 

\begin{figure*}[ht!]
    \centering
    \begin{minipage}{0.3\textwidth}
        \includegraphics[width=\linewidth]{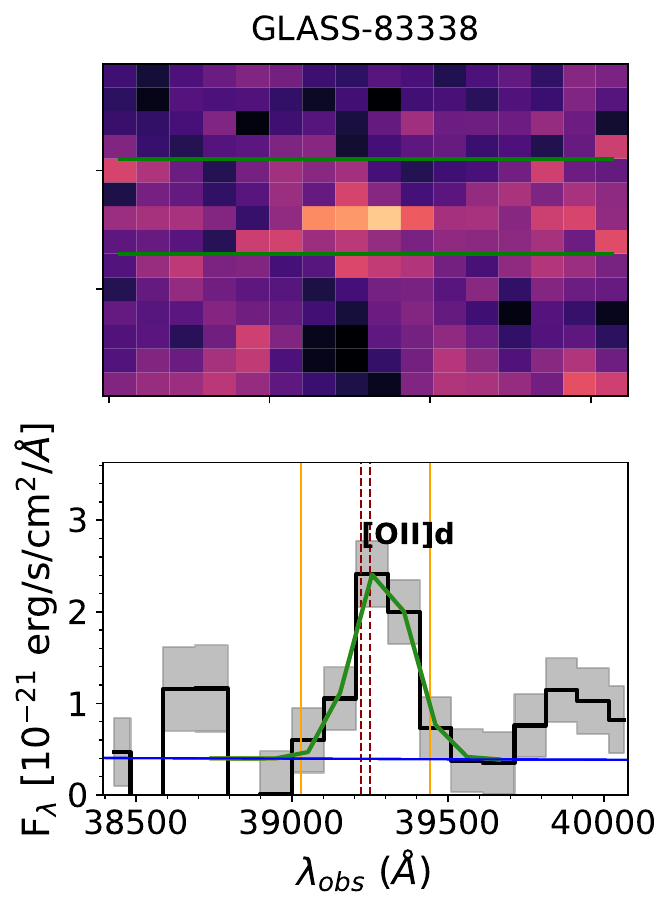}
    \end{minipage}
    \hfill
    \begin{minipage}{0.3\textwidth}
        \includegraphics[width=\linewidth]{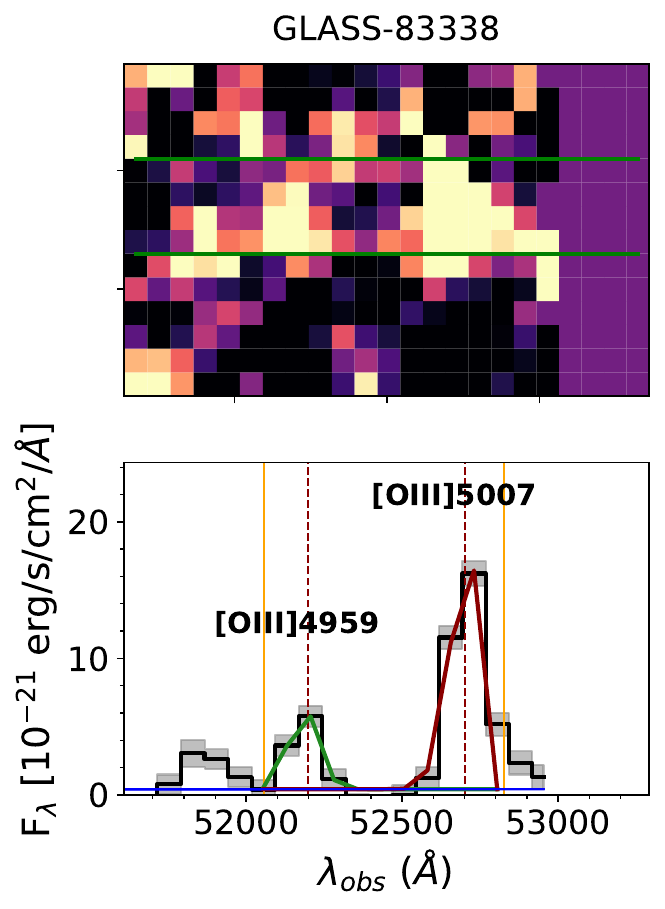}
    \end{minipage}
    \hfill
    \begin{minipage}{0.3\textwidth}
        \includegraphics[width=\linewidth]{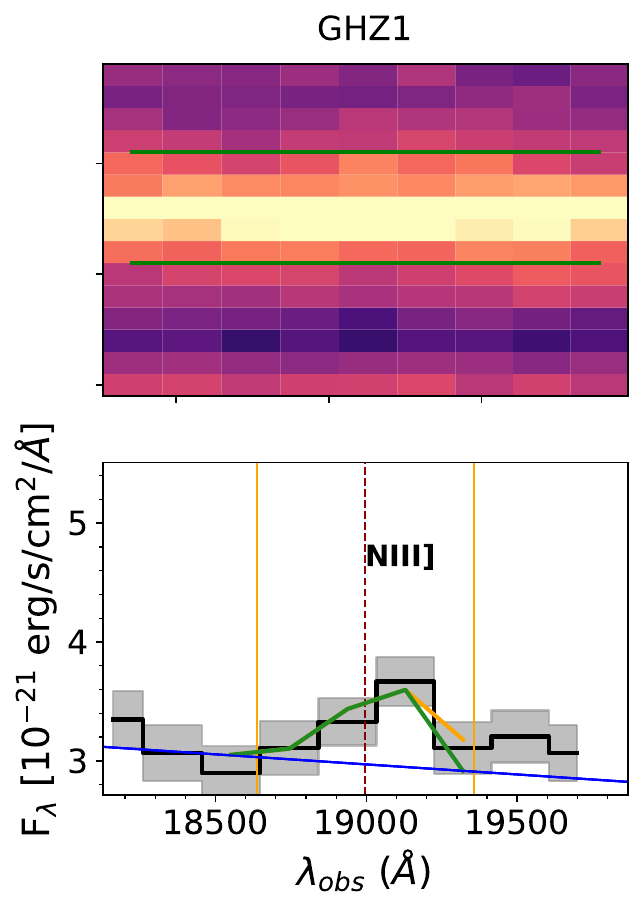}
    \end{minipage}
    
    \medskip

    \begin{minipage}{0.3\textwidth}
        \includegraphics[width=\linewidth]{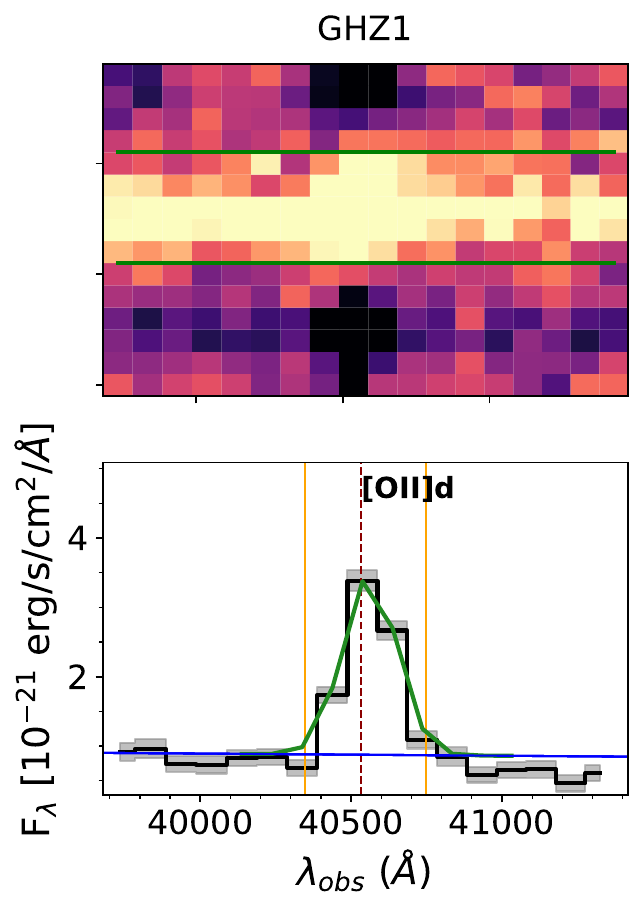}
    \end{minipage}
    \hfill
    \begin{minipage}{0.3\textwidth}
        \includegraphics[width=\linewidth]{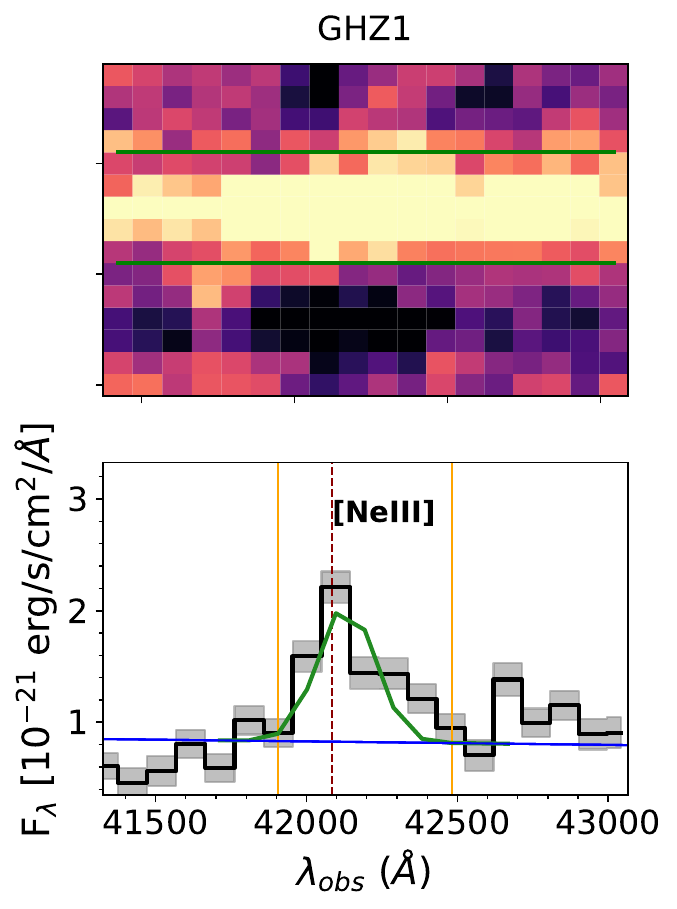}
    \end{minipage}
    \hfill
    \begin{minipage}{0.3\textwidth}
        \includegraphics[width=\linewidth]{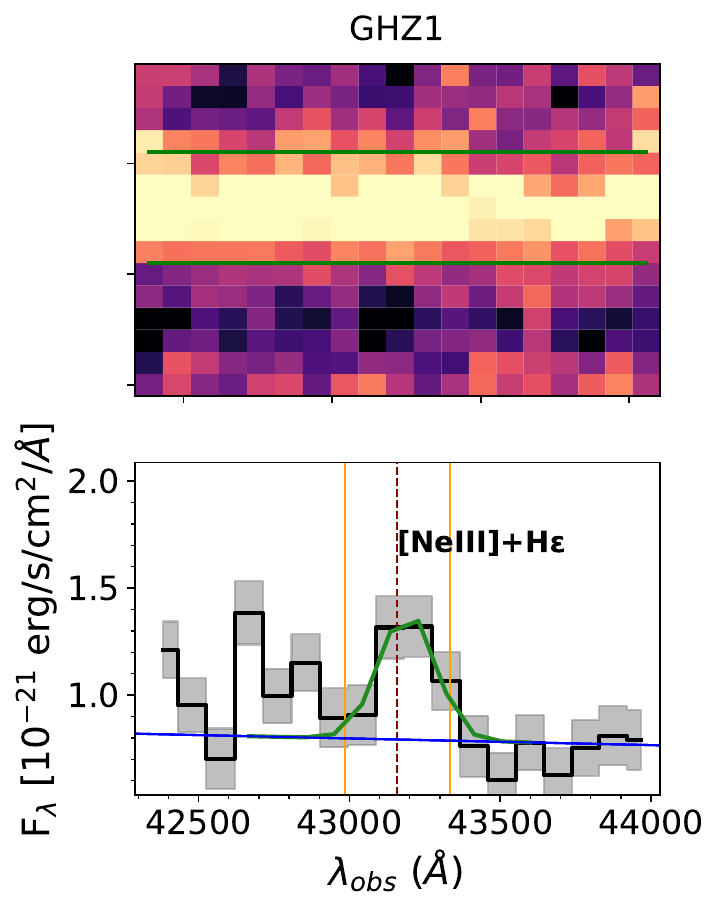}
    \end{minipage}

    \medskip

    \begin{minipage}{0.3\textwidth}
        \includegraphics[width=\linewidth]{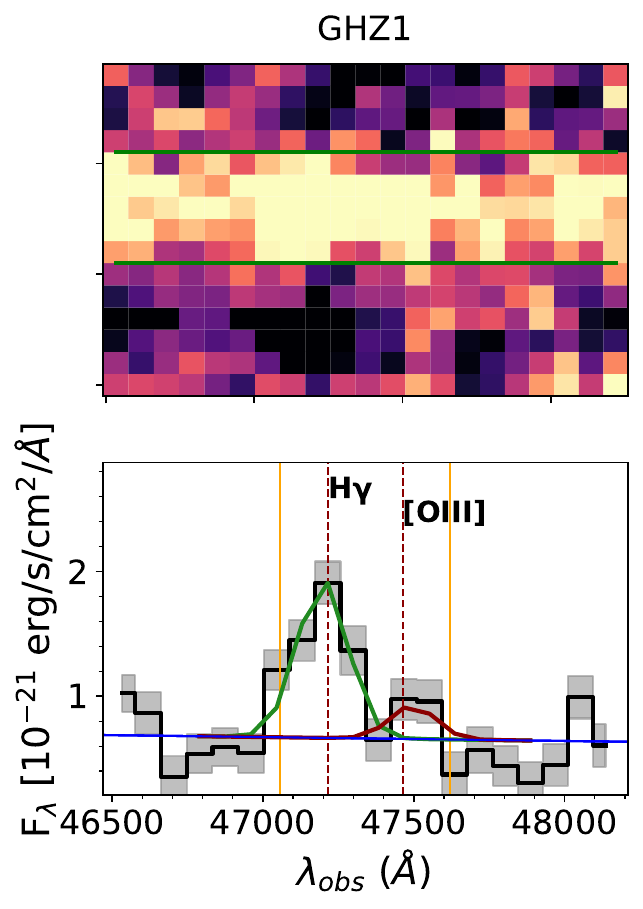}
    \end{minipage}
    \hfill
    \begin{minipage}{0.3\textwidth}
        \includegraphics[width=\linewidth]{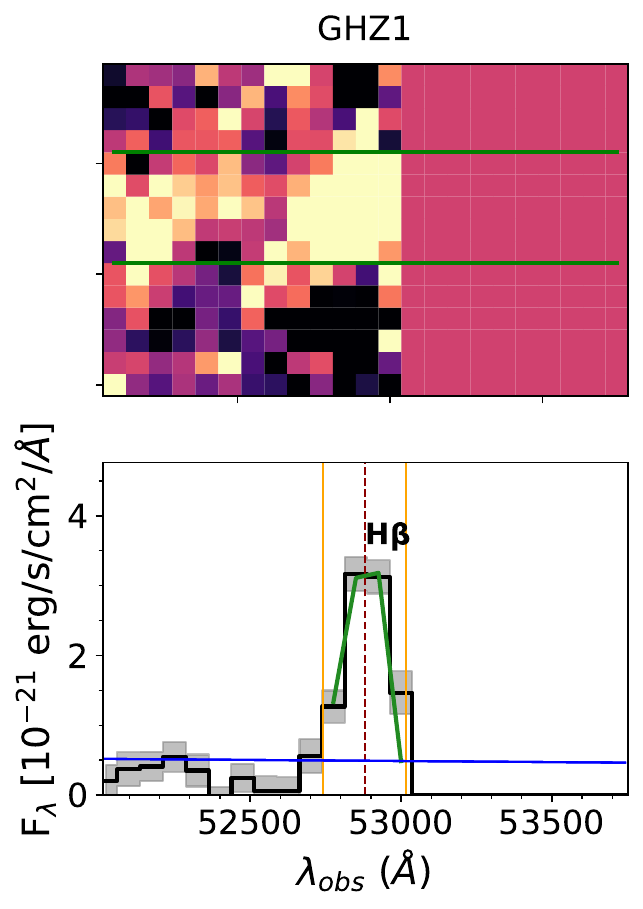}
    \end{minipage}
    \hfill
    \begin{minipage}{0.3\textwidth}
        \includegraphics[width=\linewidth]{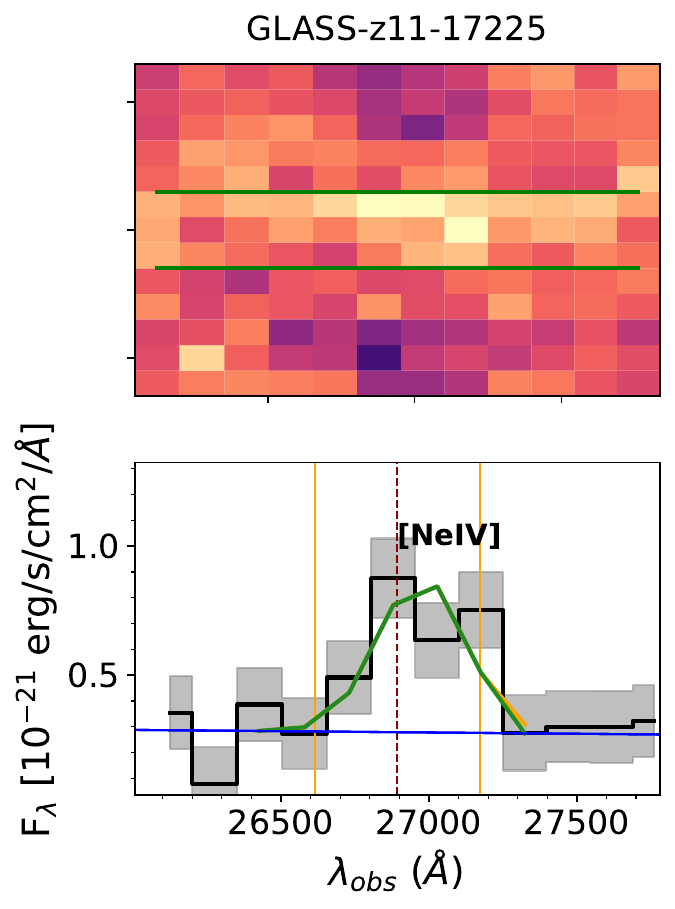}
    \end{minipage}
    
    \caption{Best-fit Gaussian models for all the significant lines. The vertical orange lines highlight the spectral region where we evaluate the S/N of the features from direct integration. As in Fig.~\ref{fig:spectra}, horizontal green lines show the customized extraction apertures. Fluxes are corrected for the lensing magnification coefficients reported in Table \ref{tab:summary_data}.}
    \label{fig:emissions}
\end{figure*}

\begin{figure*}[ht!]
    \ContinuedFloat

    \centering
    \begin{minipage}{0.3\textwidth}
        \includegraphics[width=\linewidth]{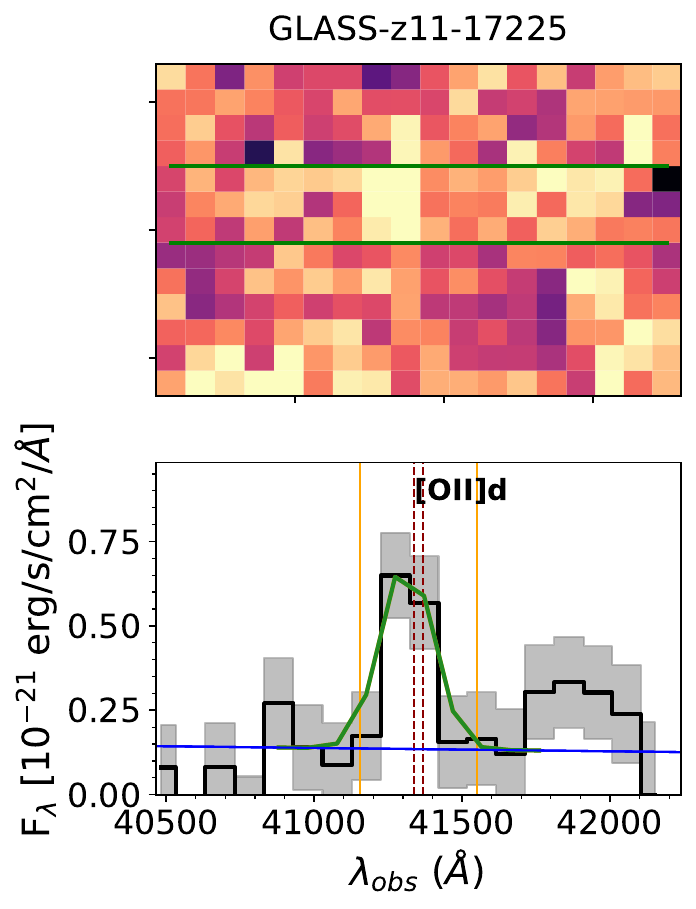}
    \end{minipage}
    \hfill
    \begin{minipage}{0.3\textwidth}
        \includegraphics[width=\linewidth]{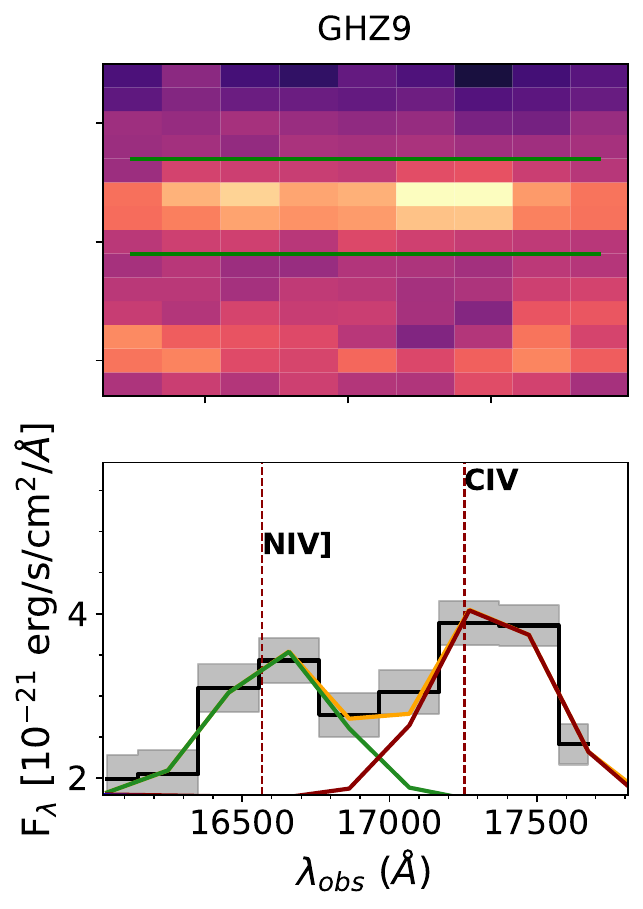}
    \end{minipage}
    \hfill
    \begin{minipage}{0.3\textwidth}
        \includegraphics[width=\linewidth]{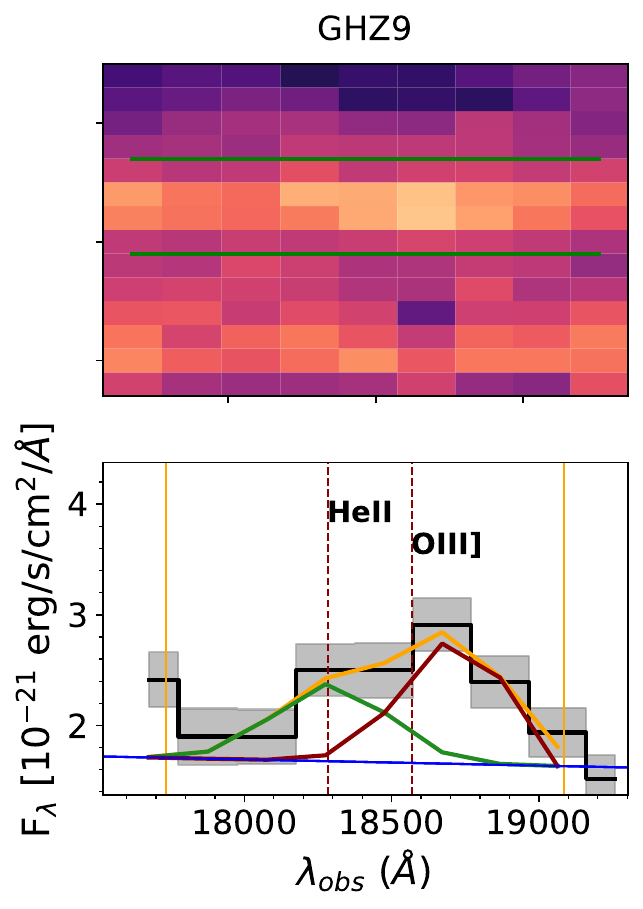}
    \end{minipage}

    \medskip
    
    \begin{minipage}{0.3\textwidth}
        \includegraphics[width=\linewidth]{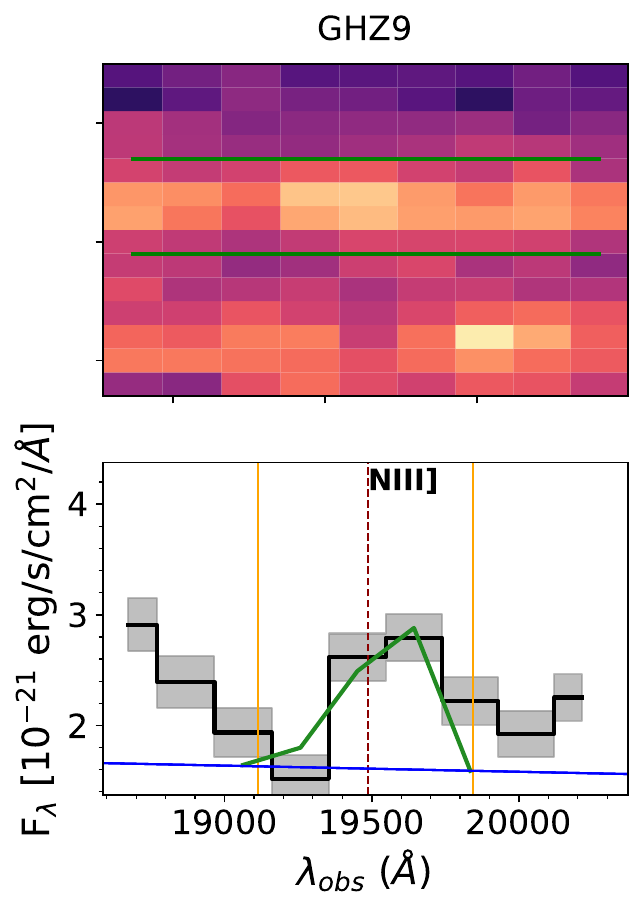}
    \end{minipage}
    \hfill
    \begin{minipage}{0.3\textwidth}
        \includegraphics[width=\linewidth]{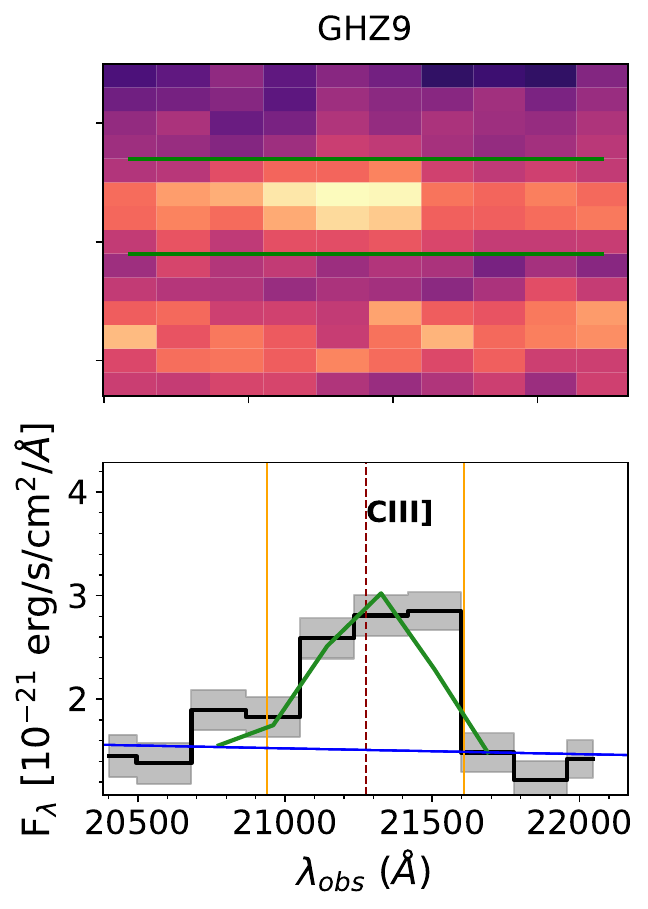}
    \end{minipage}
    \hfill
    \begin{minipage}{0.3\textwidth}
        \includegraphics[width=\linewidth]{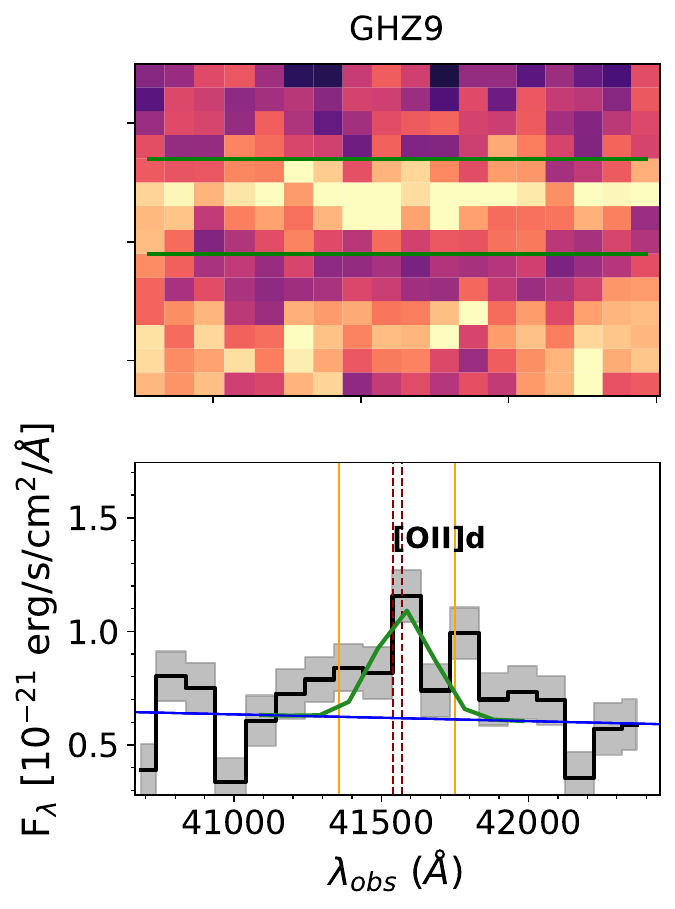}
    \end{minipage}

    \centering
    \begin{minipage}{0.3\textwidth}
        \includegraphics[width=\linewidth]{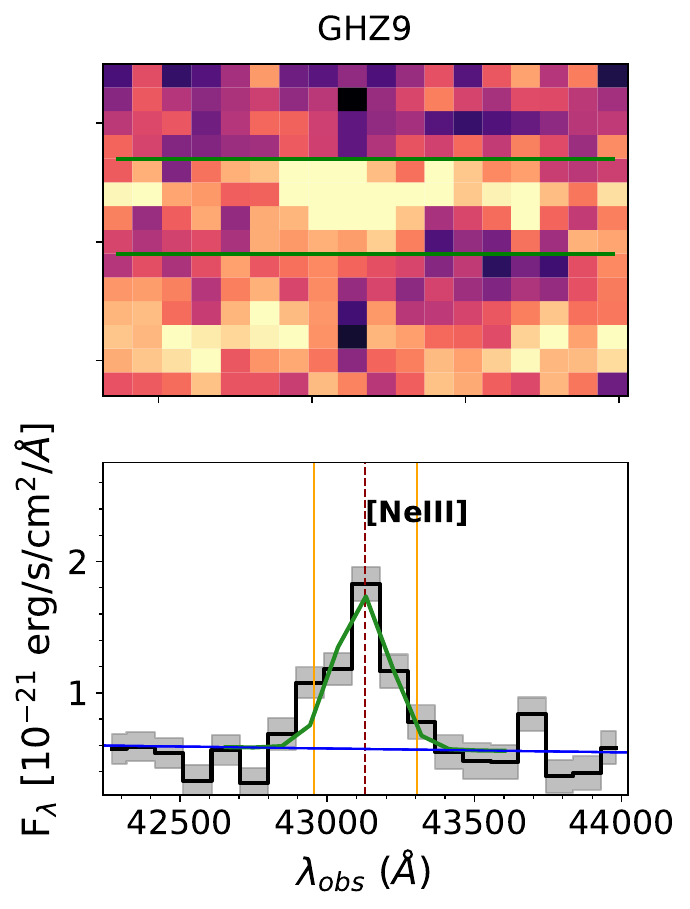}
    \end{minipage}
    \hfill
    \begin{minipage}{0.3\textwidth}
        \includegraphics[width=\linewidth]{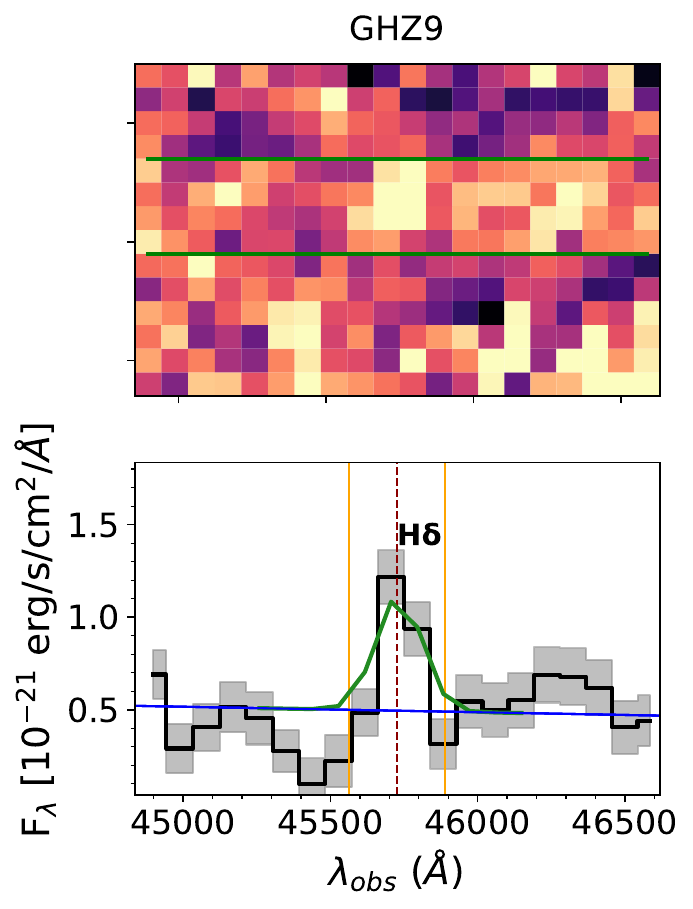}
    \end{minipage}
    \hfill
    \begin{minipage}{0.3\textwidth}
        \includegraphics[width=\linewidth]{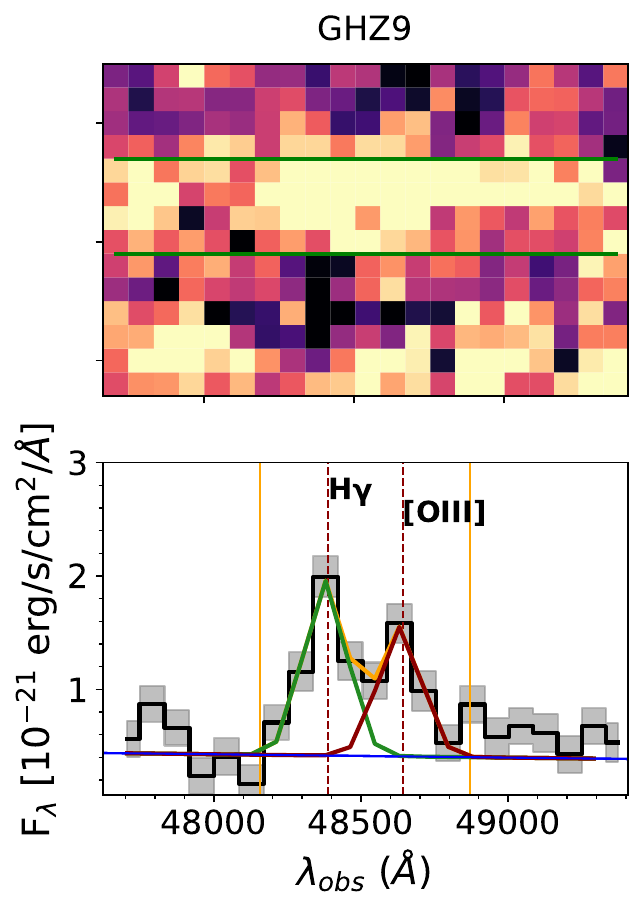}
    \end{minipage}

    \caption{continued.}
    \label{fig:emissions_continued1}
\end{figure*}

\begin{figure*}[ht!]
    \ContinuedFloat
    
    \centering
    \begin{minipage}{0.3\textwidth}
        \includegraphics[width=\linewidth]{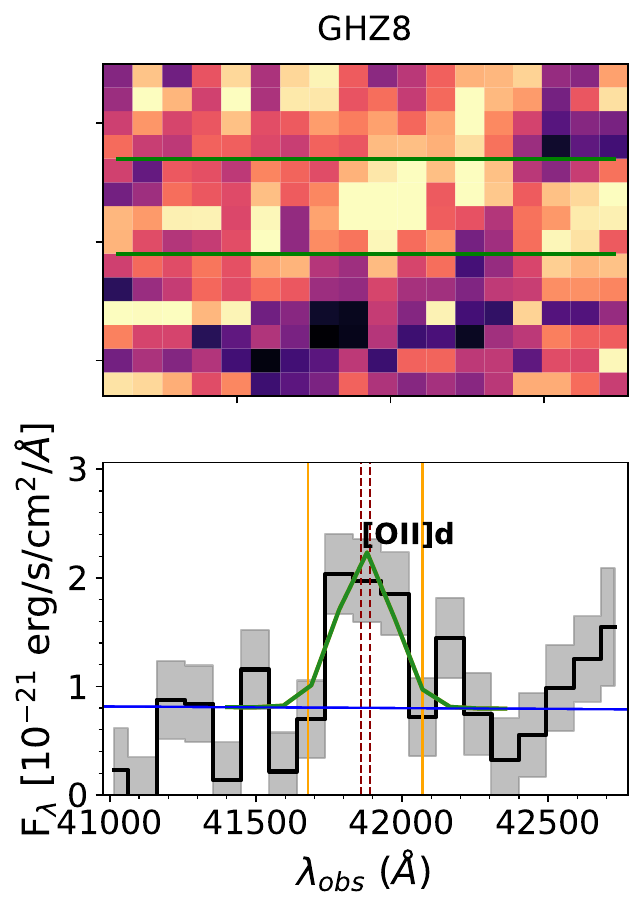}
    \end{minipage}
    \hfill
    \begin{minipage}{0.3\textwidth}
        \includegraphics[width=\linewidth]{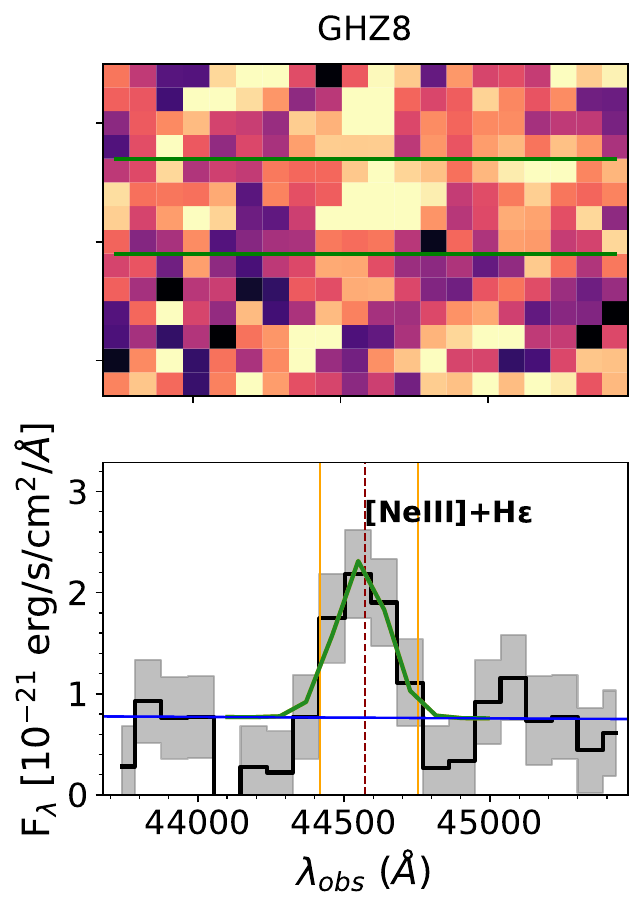}
    \end{minipage}
    \hfill
    \begin{minipage}{0.3\textwidth}
        \includegraphics[width=\linewidth]{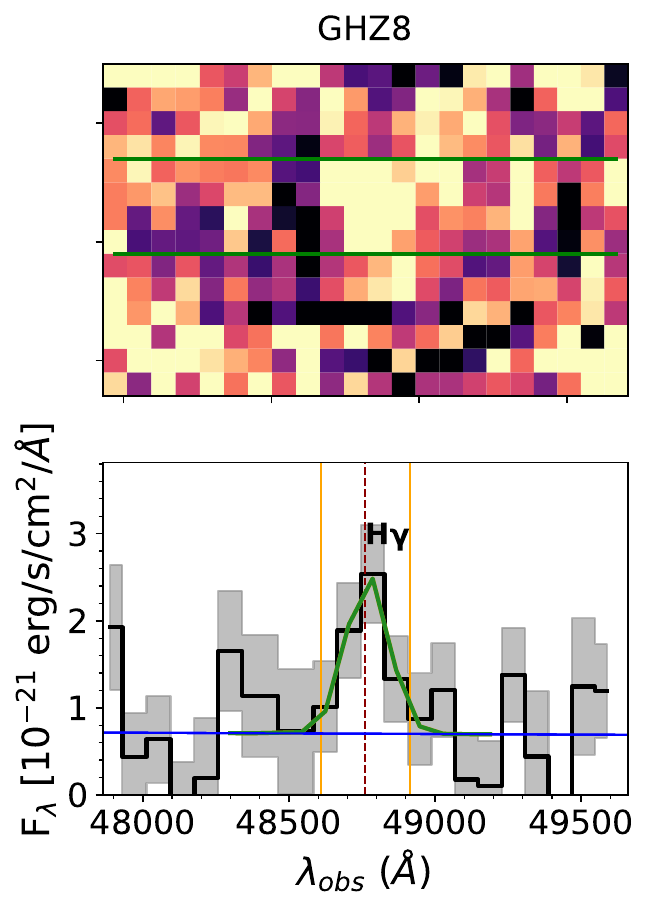}
    \end{minipage}

    \medskip
    
    \begin{minipage}{0.3\textwidth}
        \includegraphics[width=\linewidth]{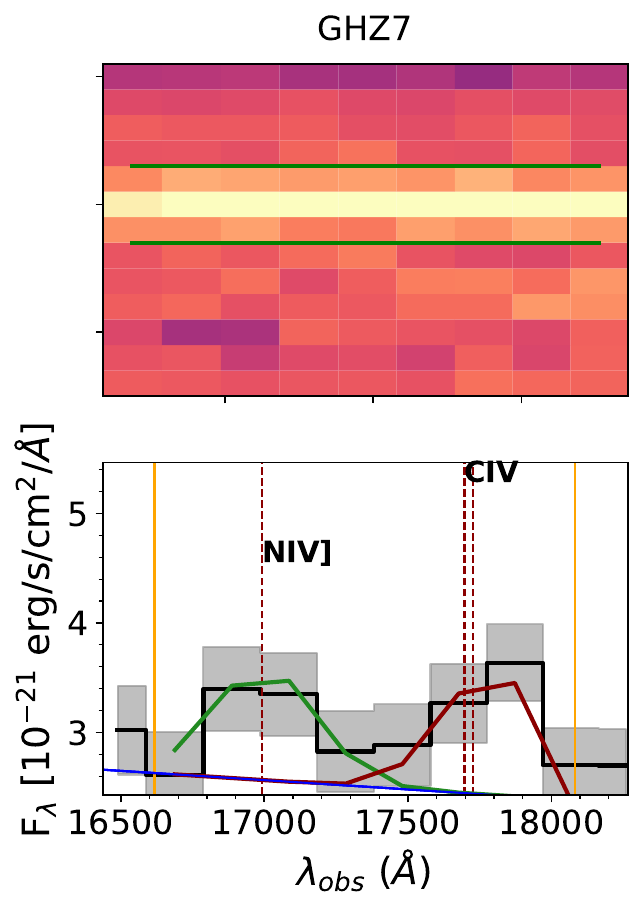}
    \end{minipage}
    \hfill
    \begin{minipage}{0.3\textwidth}
        \includegraphics[width=\linewidth]{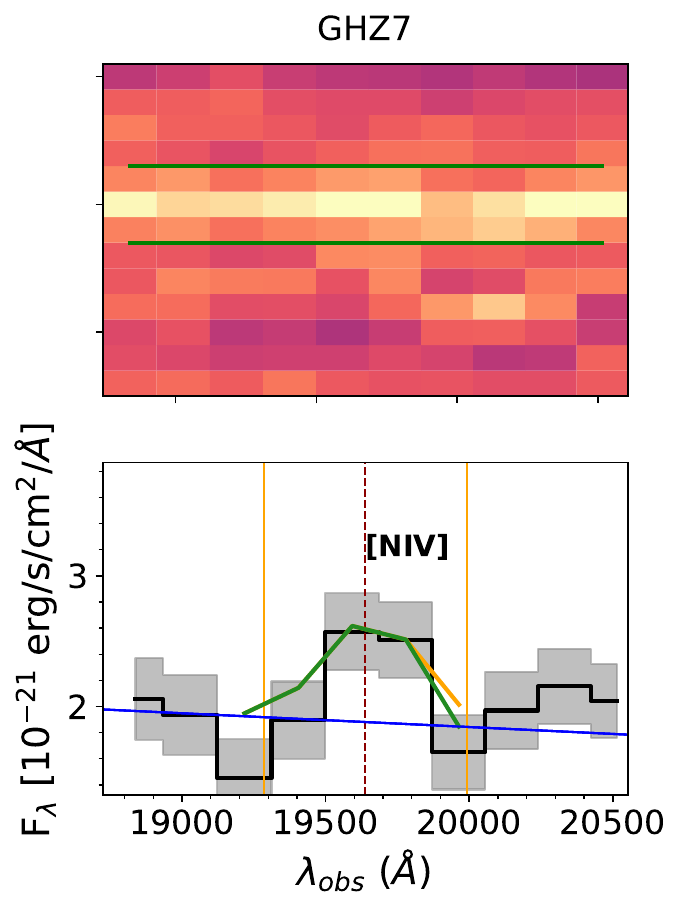}
    \end{minipage}
    \hfill
    \begin{minipage}{0.3\textwidth}
        \includegraphics[width=\linewidth]{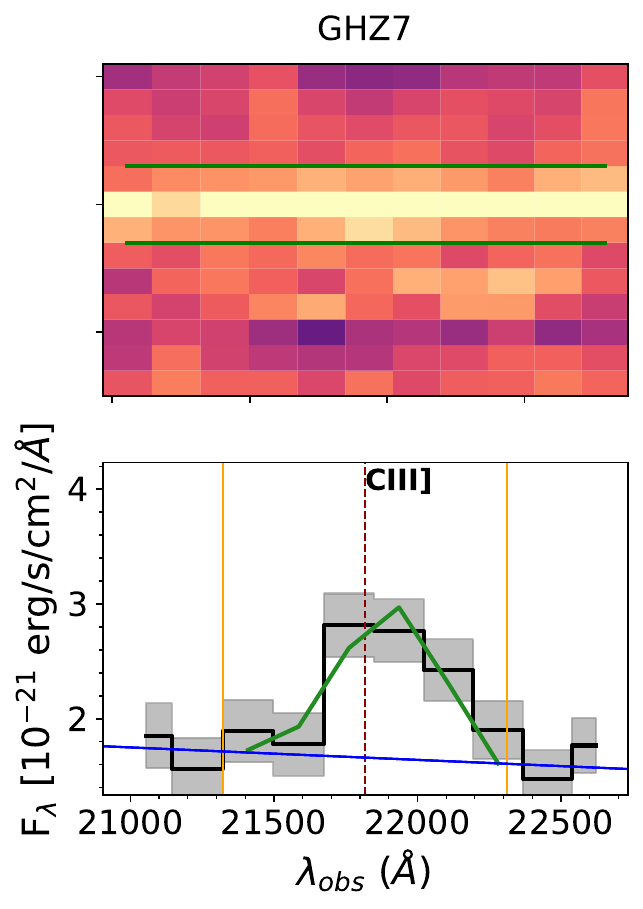}
    \end{minipage}

    \centering
    \begin{minipage}{0.3\textwidth}
        \includegraphics[width=\linewidth]{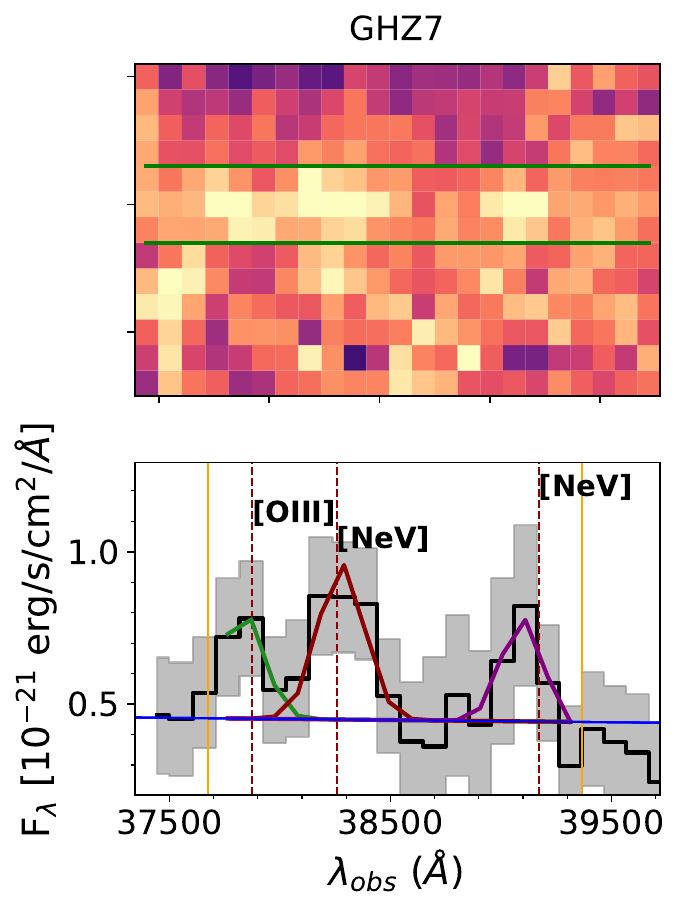}
    \end{minipage}
    \hfill
    \begin{minipage}{0.3\textwidth}
        \includegraphics[width=\linewidth]{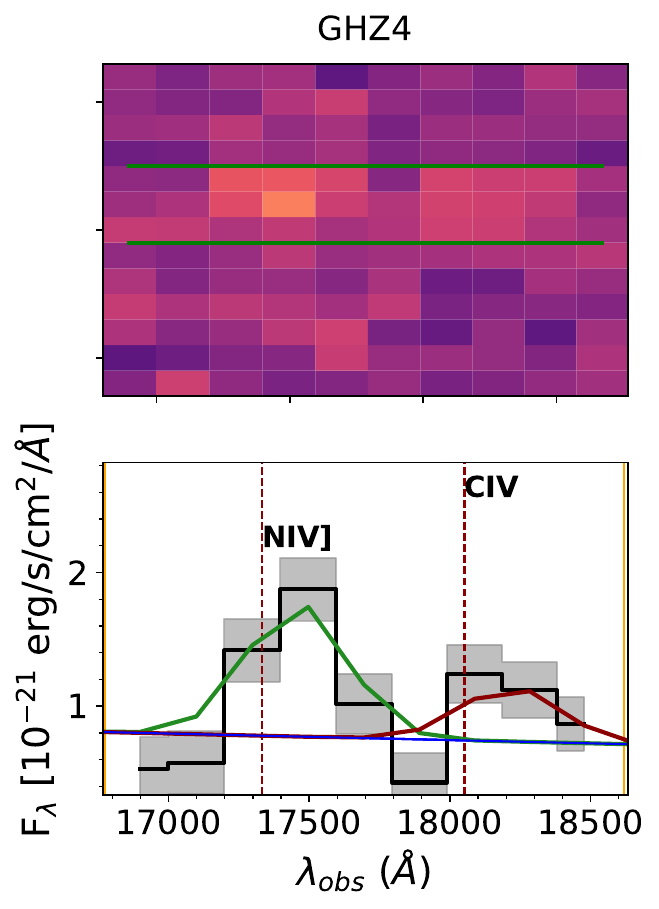}
    \end{minipage}
    \hfill
    \begin{minipage}{0.3\textwidth}
        \includegraphics[width=\linewidth]{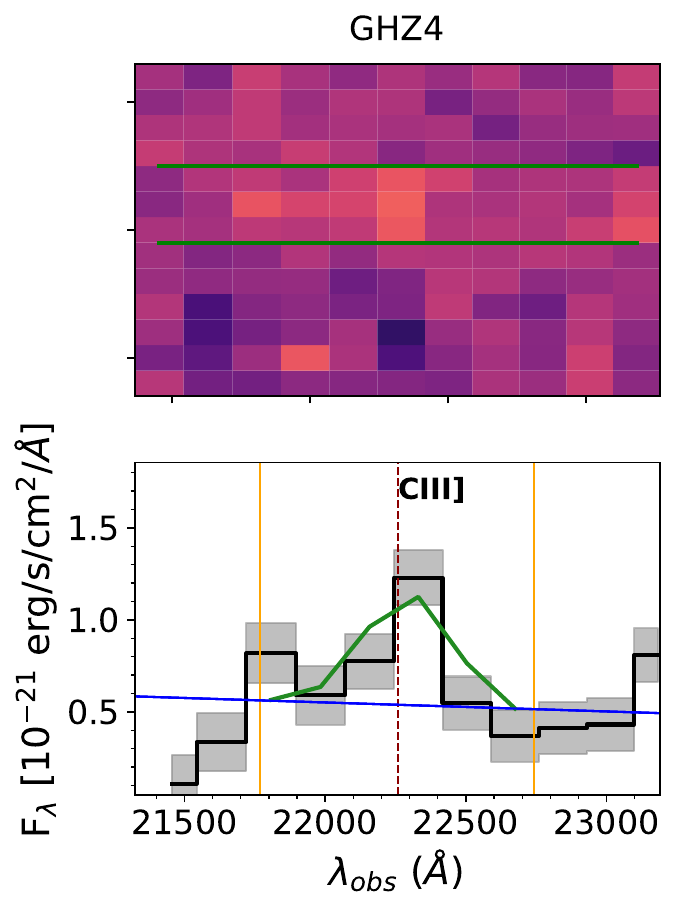}
    \end{minipage}
    
    \caption{continued.}
    \label{fig:emissions_continued2}
\end{figure*}

\begin{figure*}[ht!]
    \ContinuedFloat

    \centering
    \begin{minipage}{0.3\textwidth}
        \includegraphics[width=\linewidth]{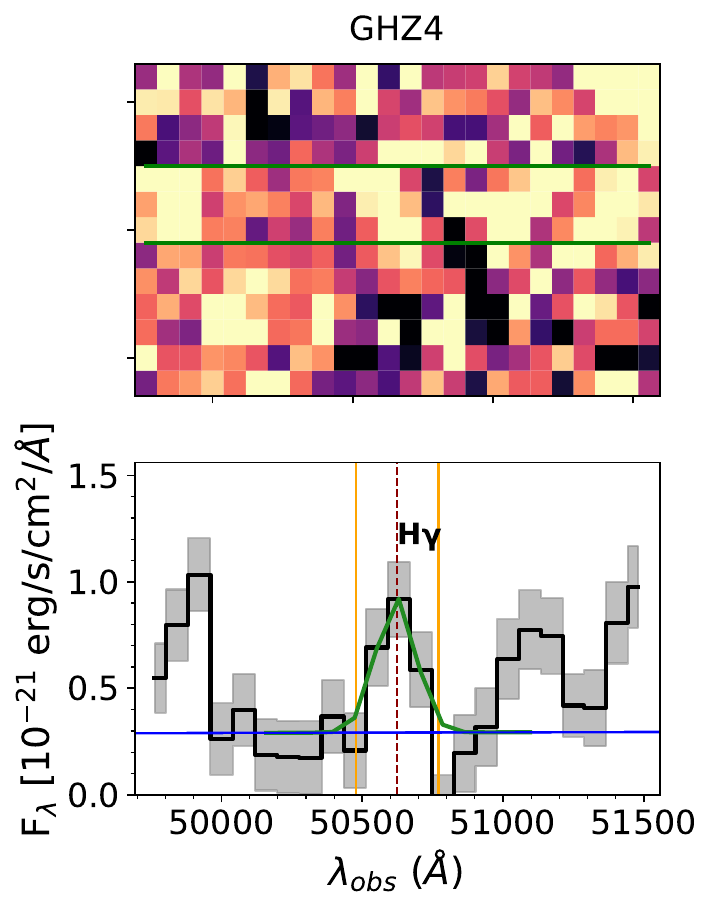}
    \end{minipage}
    \caption{continued.}
    \label{fig:emissions_continued3}
    
\end{figure*}

\end{document}